\documentclass[12pt]{article}
\newcommand{\rotpicsmall}[1]{\unitlength1cm

\begin{picture}(6.5,5.5)
\put(0,5.55){\includegraphics{#1}}
\end{picture}
}

\newcommand{\mat}[1]{\left( \begin{array}{cc} #1 \end{array} \right)}

\newcommand{\vect}[1]{\left( \! \begin{array}{c} #1 \end{array} \! \right)}

\newcommand{\thetabar}{\bar{\theta}}
\newcommand{\ri}{\frac{1}{\rho}}

\newcommand{\cm}{\cos \mu\,}
\newcommand{\sm}{\sin \mu\,}
\newcommand{\cn}{\cos \nu\,}
\newcommand{\sn}{\sin \nu\,}
\newcommand{\ra}{\rangle}
\newcommand{\la}{\langle}
\newcommand{\ua}{\uparrow}
\newcommand{\da}{\downarrow}

\newcommand{\half}{\frac{1}{2}}

\newcommand {\p}   {$^{\prime}$}
\newcommand {\ncl} {non-contractible loop}

\newcommand {\ncls}{non-contractible loops}
\newcommand {\ncs} {non-contractible sphere}

\newcommand {\art} {article}
\newcommand {\YMth}  {Yang-Mills theory}
\newcommand {\YMths} {Yang-Mills theories}
\newcommand {\YMH}   {Yang-Mills-Higgs}
\newcommand {\YMHth} {Yang-Mills-Higgs theory}

\newcommand {\YMparH} {Yang-Mills\-(\mbox{-Higgs})}
\newcommand {\YMparHth} {Yang-Mills\-(\mbox{-Higgs}) theory}

\newcommand {\cs} {configuration space}
\newcommand {\ew} {electroweak}

\newcommand {\Sstar} {S$^{\star }$}

\newcommand {\ewsm} {electroweak standard model}
\def\id{{\rm 1\kern-.12em
\rule{0.3pt}{1.5ex}\raisebox{0.0ex}{\rule{0.1em}{0.3pt}}}}

\def\C{{\rm\kern.24em
   \vrule width.02em
       height1.4ex depth-.05ex
   \kern-.26em C}}

\def\R{{\rm I\kern-.25em R}}

\def\Z{\mbox{\sf Z\hspace{-1.8ex} \sf Z}}   

\def\N{{\rm I\kern-.23em N}}

\newcommand{\pdq}[2]{\frac{\partial #1}{\partial #2}}

\newcommand{\beq}{\begin{equation}}
\newcommand{\eeq}{\end{equation}}
\newcommand{\bea}{\begin{eqnarray}}
\newcommand{\eea}{\end{eqnarray}}

\renewcommand{\phi}{\varphi}

\begin{document}

\begin{titlepage}
\hspace*{\fill} hep-th/9702023
\newline
\hspace*{\fill} KA--TP--01--1997 
\begin{center}
\vspace{3\baselineskip}
{\Large \bf A global anomaly from the $Z$-string}\\
\vspace{1\baselineskip}
{\large F. R. Klinkhamer and C. Rupp} \\
\vspace{1\baselineskip}
 Institut f{\"u}r Theoretische Physik\\ Universit{\"a}t Karlsruhe\\
 76128 Karlsruhe\\ Germany \\
\vspace{3\baselineskip}
{\bf Abstract} \\
\end{center}
{\small
\noindent 
The response of isodoublet fermions to  classical
backgrounds of essentially 2-dimensional boson fields in
$SU(2)$ \YMHth ~is investigated.
In particular, the spectral flow of Dirac eigenvalues
is calculated for a \ncs ~of configurations passing through the vacuum and the
$Z$-string (the embedded vortex solution). Also, a non-vanishing
Berry phase is established for adiabatic transport ``around'' the $Z$-string.
These results imply the existence of a new type of global (non-perturbative)
gauge anomaly in $SU(2)$ \YMH ~quantum field theory with a single
doublet of left-handed fermions. Possible extensions to other chiral
gauge field theories are also discussed.
}
\vspace{2\baselineskip}
\begin{tabbing}
PACS numbers \= : \= 11.27.+d, 03.65.Pm, 03.65.Bz, 11.15.-q\\
keywords     \> : \> classical solution, spectral flow, Berry phase, anomaly
\end{tabbing}
\vfill
to be published in Nucl. Phys. {\bf B }

\end{titlepage}

\section{Introduction}

A special class of static classical field configurations in 3+1 dimensional
\YMHth ~consists of those configurations that
are constant in a fixed spatial direction. The \cs ~of these essentially
2-dimensional fields has non-trivial structure.
For $SU(2)$ \YMHth ~in particular, a \ncs ~has been constructed \cite{KO} with 
the classical vacuum at the bottom and 
the embedded Nielsen-Olesen vortex solution or $Z$-string \cite{NO,N} at the top.
In this paper we calculate for isodoublet fermions the spectral flow of the Dirac Hamiltonian
when the bosonic background fields range over
this \ncs. Also, we show that adiabatic transport of the first-quantized fermion states
along a curve of these bosonic background fields close to and around the $Z$-string
gives rise to a non-vanishing Berry phase \cite{B}, with the value $\pi$.

This curve close around the $Z$-string can be pulled down into the vacuum. 
The result is a \ncl ~of vacuum configurations of the classical boson fields.
The second-quantized fermionic vacuum then exhibits
a M{\"o}bius bundle structure over this particular gauge orbit.
This corresponds to a global (non-perturbative)
gauge anomaly of the chiral gauge field theory, which is analogous to,
but not the same as, the Witten anomaly \cite{W}.
The reasoning behind these last steps has been reviewed
in \cite{NAG}, to which the reader is referred for further
details. The main goal in this paper is to establish, at the first-quantized level,
the spectral flow and the Berry phase. This is done, in detail, for chiral $SU(2)$
\YMHth ~with a single doublet of left-handed fermions. 
The obvious question how this extends to   
other chiral gauge theories will be discussed briefly at the end of this article.
At this point, we should mention that the results of the present paper
are discussed in a  $3+1$ dimensional context, but they
may also be relevant to lower dimensional theories.

The outline of this paper is as follows. In Section 2, the \ncs ~of
classical $SU(2)$ gauge and Higgs fields is presented. 
Actually, three related sets of configurations are constructed, 
with the topology of a balloon, a sphere and a disc.
In Section 3, the spectral flow of Dirac eigenvalues over the ``balloon''
is calculated numerically. In Section 4, a Berry phase factor $-1$
is established for non-trivial loops on the ``balloon'' and on the ``sphere''.
In Section 5, this Berry phase factor is pulled down over
the ``disc'' to a \ncl ~in the vacuum. The results from Sections 3 to 5 are
valid at the first-quantized level. In Section 6, finally, the resulting
global gauge anomaly of the chiral $SU(2)$ \YMH ~quantum field theory
and possible extensions
to other chiral gauge field theories are discussed. This last Section
is reasonably self-contained. Four Appendices give some details
of the calculation, alternative derivations and miscellaneous results.
Natural units ($\hbar$ $=$ $c$ $=$ $1$) are used throughout.

\section{Classical background fields}

As mentioned in the introduction,
our main focus is on $SU(2)$ \YMHth, which corresponds to the \ewsm ~in
the limit of vanishing weak mixing angle $\theta_w$.
The usual Minkowski space-time is modified to have for the spatial coordinate
$x^3$  a very large, but finite, range $\ell_3$.  
The boson fields of the theory are the $SU(2)$
Gauge field $W_{\mu}(x)$ and the complex Higgs doublet field $\Phi(x)$.
The energy density for static (time-independent) classical boson fields,
in the temporal gauge $W_0=0$, is given by
\beq
 e_{\, \rm b} =
     \frac{1}{4 \,g^{2}}\; \left( W_{m n}^{a} \right)^{2} +
    |D_{m} \Phi|^{2} +
     \lambda \left( |\Phi|^{2} - \frac{v^{2}}{2} \right)^{2} \, ,
\label{eq:e}
\eeq
with the covariant derivative and field strength 
\begin{eqnarray*}
D_{m}\Phi&\equiv&
\left( \partial_{m} + W_{m} \right) \Phi \equiv
\left( \partial_{m} + W_{m}^{a}\; \frac{\tau_{a}}{2 \,i}\right) \Phi \, , \\
W_{mn} &\equiv& 
  \partial _{m} W_{n} - \partial _{n} W_{m} + [W_{m},W_{n}] \equiv
  W_{mn}^{a}\: \frac{\tau_{a}}{2\,i}  \, ,
\label{eq:definitions}
\end{eqnarray*}
where the indices $m$, $n$ and $a$ run over the values $1,\, 2,\,3$, and
$\tau_{a}$ are the Pauli matrices
\[
    \tau_1 \equiv \left( \begin{array}{cc} 0 & 1 \\ 1 & 0 \end{array} \right) \, ,
    \qquad
    \tau_2 \equiv \left( \begin{array}{cc} 0 & -i \\ i & 0 \end{array} \right) \, ,
    \qquad
    \tau_3 \equiv \left( \begin{array}{cc} 1 & 0 \\ 0 & -1 \end{array} \right) \, .
\]
The semiclassical masses of the three degenerate $W$ vector 
bosons\footnote{The $Z_{\mu}$ field in the \ew ~context for $\theta_w=0$ is simply
defined as the $a=3$ component of the $SU(2)$ gauge fields $W^a_{\mu}$.}
and the single Higgs scalar boson are
$M_{W}=\frac{1}{2}\, g\, v$ and $M_{H} = \sqrt{2 \lambda} \, v$, respectively.

In this \art ~we consider static
field configurations that are constant in the $x^3$ direction
and have the corresponding component of the gauge field vanishing.
Effectively, one has a 2-dimensional theory with energy density
(\ref{eq:e}), where the spatial indices take only the values $m=1,\,2$.
These two spatial dimensions can also be described by the cylindrical
coordinates $\rho$ and $\phi$, defined in terms of the cartesian coordinates by
\[(x^{1},\,x^{2},\,x^{3}) \equiv (\rho \cos \phi,\,  \rho \sin \phi,\, z)\, .\]
Furthermore, we consider only field configurations
with {\em finite\/} string tension (energy per unit of length in the $x^3$ direction).
Of course, fields with finite, non-zero string tension $\sigma$ have very large
total energy $E_{\, \rm b}= \sigma \,\ell_3$.
The total energy $E_{\, \rm b}$ of these essentially 2-dimensional fields can be
zero only if the string tension $\sigma$ vanishes exactly.

 A two parameter family of such SU(2) gauge and Higgs fields
 has been given \cite{KO} that represents a non-contractible sphere
 (NCS) in configuration space. The field configurations
 $W \equiv \sum_{m=1,2} \,W_m {\rm d}x^m$ and $\Phi$ are defined as follows
 \bea
  W    &=& -f \,  {\rm d}U \, U^{-1} \, ,\nonumber\\ [0.2cm]
  \Phi &=& \frac{v}{\sqrt{2}} \, h \, U \vect{0\\1},
 \label{felddef}
 \eea
 with the $SU(2)$ valued functions
 \bea
  U(\mu,\nu,\phi) &=& \Omega \, M         \, , \nonumber\\[0.2cm]
  M(\mu,\nu,\phi) &=&
\left( \begin{array}{l}
        \sm\\ \cm\sn \\ \cm\cn\sin\phi \\ \cm\cn\cos\phi
       \end{array}
\right) \cdot
\vect{-i\tau_1\\-i\tau_2\\-i\tau_3\\ \id_2}\,, \nonumber\\[0.2cm]
 \Omega(\mu,\nu) &=& M(\mu,\nu,0)^{-1} = \sm \, i\tau_1 + \cm\sn
 \, i\tau_2 + \cm\cn \, \id_2\, ,
\label{matrizen}
 \eea
where $\mu, \nu\in [-\pi/2,\pi/2]$ parametrize the sphere and $\phi\in
[0,2\pi]$ is the azimuthal  coordinate.
The map $U(\mu,\nu,\phi)$  is topologically non-trivial
($S_2 \times S_1 \rightarrow S_3$, with winding number $n=1$).
The field configurations (\ref{felddef}) at the top of the sphere ($\mu=\nu=0$)
correspond to the $Z$-string,
which is the Nielsen-Olesen vortex \cite {NO} of the $U(1)$
Abelian Higgs model embedded \cite{N} in the $SU(2)$ \YMHth.
The profile functions $f=f(\rho)$ and  $h=h(\rho)$
solve the reduced field equations (for the ansatz at $\mu=\nu=0$)
with boundary conditions
\beq
f(0)= h(0)= 0 \; , \quad f(\infty)= h(\infty)= 1 \; .
\label{eq:NObcs}
\eeq
These conditions guarantee that the fields (\ref{felddef}) are regular 
at the origin and pure gauge at infinity.

\begin{figure}
\unitlength1cm
\begin{center}

\begin{picture}(15,8)
\put(0,8.5){\includegraphics{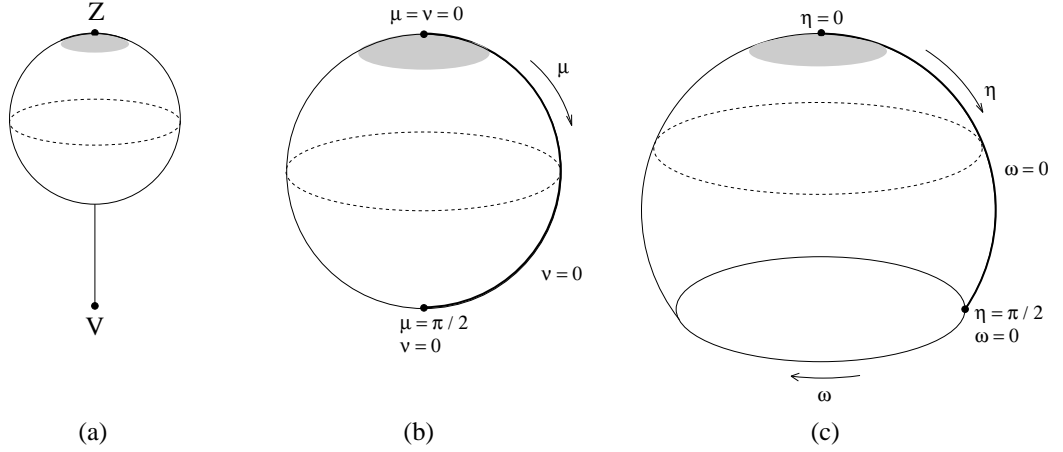}}
\end{picture}
\end{center}
\caption{Sketch of (a) the $Z$-NCS\p, (b) the $Z$-NCS and (c) the $Z$-NCD.
\label{skizze}}
\end{figure}

The $W$ fields vanish identically at the bottom of the sphere 
($|\mu|=\pi/2$ or $|\nu|=\pi/2$), but the Higgs field is still
different from its vacuum value.
This Higgs configuration can be connected to the vacuum V by an additional
line segment on which the function $h(\rho)$ is interpolated to the
constant $1$. Taking the total range for $\mu$ and $\nu$ to be $[-\pi,\pi]$,
we define for $[\mu\nu] \equiv \max \left( |\mu|, |\nu|\right) \geq \pi/2$
\begin{eqnarray}
W    &=&0 \, , \nonumber\\
\Phi &=& \frac{v}{\sqrt{2}} \,
         (\,1-(1-h) \, \sin [\mu\nu]\,) \vect{0\\1}. \label{schnur}
\label{felddef2}
\end{eqnarray}
The family of field configurations (\ref{felddef})--(\ref{schnur}) has
the topology of a balloon (Fig. \ref{skizze}a), and, with a slight
abuse of terminology, we call this whole set of configurations the
$Z$-NCS\p ~(the prime is to remind us of the balloon string).

The fermion calculations of the next Section will be performed,
for purely technical reasons,
on the background fields (\ref{felddef}, \ref{matrizen}) with
the matrix $\Omega$ eliminated by a global gauge transformation. The
``balloon'' then has a hole, because the rim of the square $|\mu|,|\nu| \le
\pi/2$ is no longer mapped into a single configuration. Making an
appropriate global gauge transformation of the fields (\ref{felddef2}),
one can say that the balloon (Fig. \ref{skizze}a) will be temporarily replaced
by a ``flask''.

The energy functional over the $Z$-NCS\p ~fields (\ref{felddef}, \ref{matrizen})
contains, as shown in \cite{KO}, the factor
$\cos^2\mu\cos^2\nu$ with a positive coefficient, so that the manifest
maximum energy configuration corresponds to the $Z$-string at
$\mu=\nu=0$. This was the reason to introduce separately the ``string''
of configurations (\ref{schnur}) that ties the ``balloon'' to the
vacuum. A more unified set of configurations,
for $\mu, \nu\in [-\pi/2,\pi/2]$,  is the following:
\bea
 W &=& -f\,  {\rm d}U \, U^{-1}\, , \nonumber\\
 \Phi &=& \frac{v}{\sqrt{2}} \, \Omega \, \left[ \,h\, M\, + \, (1-h)\,
 \left( \sm (-i\tau_1)\, + \, \cm\sn (-i\tau_2) \right) \,\right] \,
   \vect{0\\1},
\label{ZNCSfields}
\eea
with $U$, $M$ and $\Omega$ as given in (\ref{matrizen}).
The family of field configurations (\ref{ZNCSfields}) has the topology of a
sphere (Fig. \ref{skizze}b), with the $Z$-string solution at $\mu=\nu=0$
and the vacuum V at $[\mu\nu]=\pi/2$, and we call this set of
configurations the $Z$-NCS.  Clearly, the fields of the $Z$-NCS\p ~and
the $Z$-NCS can be interpolated continuously.
Remark that in both cases the local and global $SU(2)$ gauge freedom
has been eliminated by the gauge fixing conditions
\bea
W_\rho (\rho,\phi)       &=& 0 \; ,   \nonumber\\[0.2cm]
\Phi(\rho=\infty,\phi=0) &=& \frac{v}{\sqrt{2}} \, \vect{0\\1}  \, .
\label{radialgauge}
\eea

Finally, we introduce a set of configurations with the topology of a
disc, parametrized by a radial variable $\eta \in [0,\pi/2]$ and an
angular variable $\omega \in [0,2\pi]$. The $Z$-string sits at the
middle of the disc ($\eta=0$), whereas the rim ($\eta=\pi/2$)
corresponds to a non-contractible loop 
of vacuum configurations parametrized by $\omega$ (Fig. \ref{skizze}c).
This set of configurations is called the $Z$ non-contractible disc
($Z$-NCD), even though,
properly speaking, only the vacuum loop is non-contractible.
Specifically, the field configurations are the following:
\bea
  W &=& -\sin^2 \eta \, {\rm d}U_V \, U_V^{-1}\, -\, f \cos^2\eta\,
        {\rm d}U_Z\, U_Z^{-1}
    -\, f\sin\eta\cos\eta\, \left( {\rm d}U_V\, U_Z^{-1} \, +\,
    {\rm d}U_Z \, U_V^{-1} \right) \, , \label{disc} \nonumber \\
  \Phi &=& \frac{v}{\sqrt{2}}\, \left[\, \sin\eta \, U_V \, + \, h
  \cos\eta \, U_Z \, \right] \, \vect{0\\1} \, ,
\label{ZNCDfields}
\eea
with the SU(2) matrices
\bea
U_V(\eta,\omega,\rho,\phi)&=&\Omega_D \,M_V\,,\nonumber\\[0.2cm]
U_Z(\eta,\omega,\phi)     &=&\Omega_D \,M_Z\,,\nonumber\\[0.2cm]
    \Omega_D(\eta,\omega) &=&
    \sin\eta \,(\cos\omega \, i\tau_1 -  \sin\omega \, i\tau_2)
    + \cos\eta \, \id_2 \,,                              \nonumber \\[0.2cm]
M_Z(\phi)  &=& \sin\phi\, (-i\tau_3)+\cos\phi\, \id_2 \,,\nonumber\\[0.2cm]
M_V(\omega,\rho,\phi) &=&                             
          \vect{\cos\omega + \sin^2 \frac{\omega}{2}\,
                (1+\cos\thetabar )    \\
                - \frac{1}{2} \sin\omega\, (1- \cos\thetabar ) \\
                +\sin\frac{\omega}{2} \, \sin\thetabar
                \cos\phi                                \\
                -\sin\frac{\omega}{2}\, \sin\thetabar
                \sin\phi                                }
          \cdot
          \vect{ -i\tau_1 \\ -i\tau_2 \\ -i\tau_3 \\ \id_2} \, , 
\label{ZNCDmatrices}
\eea
where $\phi \in [0,2\pi]$ is the usual
azimuthal angle and $\thetabar \in [0,\pi]$ a compactified radial
coordinate for which we take
$\thetabar = \pi \rho^2/(\rho^2+\pi)$.            
The map $U_V(\pi/2,\omega,\rho,\phi)$ is topologically non-trivial
(effectively $S_1\times
S_2 \rightarrow S_3$, with winding number $n=1$).
Note that the fields (\ref{ZNCDfields}) 
are manifestly regular at
$\rho=0$ because of the boundary conditions (\ref{eq:NObcs})
and the factor $\sin\thetabar \sim \rho^2$
multiplying $\cos\phi$ and $\sin\phi$ in $M_V$.

The fields of the $\omega=0$, $\eta\in [0,\pi/2]$ ray of the $Z$-NCD are
identical to those of the $\nu=0$, $\mu\in[0, \pi/2]$ meridian of
the $Z$-NCS, see Fig. \ref{skizze}bc .
Loosely speaking, the rest of the $Z$-NCD wraps around the $Z$-NCS.
In fact, the configurations on the caps close to the $Z$-string can be interpolated
continuously between the $Z$-NCS and the $Z$-NCD, if one identifies
$\mu = \eta \cos\omega$ and $\nu= - \eta \sin\omega$ for $\eta \sim 0$.
Note also that the circles of constant $\eta$
on the $Z$-NCD (\ref{ZNCDfields})
do not correspond to gauge orbits, except for the case of $\eta=\pi/2$
(this can be seen most easily for the Higgs field, with $\Omega_D$
eliminated and $\omega$ infinitesimal). For the $Z$-NCS (\ref{ZNCSfields})
the gauge is fixed completely
and there are no gauge orbits at all.

Having presented these three related sets of bosonic field configurations, 
we will first investigate the response of the fermions for the simplest case,
the $Z$-NCS\p.

\section{Spectral flow in the $Z$-NCS\p ~background}

Consider a single $SU(2)$ doublet of left-handed fermions $(u_L, d_L)^{\rm T}$ and
two singlets of right-handed fermions ($u_R$ and $d_R$) in the
background of the $Z$-NCS\p ~boson fields.
The fer\-mi\-on fields are combined into one doublet
\[ \Psi(x)=\vect{u(x)\\d(x)}, \]
where $u$ and $d$ are the complete, 4-component Dirac fields.         
The fermions are coupled to the Higgs  by a Yukawa term in the
Lagrangian (strictly speaking, the Lagrangian density)
\beq
 {\cal L} = -i\,\bar \Psi \, \gamma^\mu D_\mu \Psi +
g_{\,\rm Y} \left( \bar \Psi_L \,\Phi_M \, \Psi_R + 
                   \bar \Psi_R \,\Phi_M^\dagger\, \Psi_L \right)\, ,
\label{Lagrangian}
\eeq
where the covariant derivative is defined as $D_\mu \equiv \partial_\mu + W_\mu P_L$, 
with the projection operator $P_L\equiv \half(\id-\gamma_5)$,
and $\Phi_M$ is the Higgs field written as a matrix
\[ \Phi_M (x)= \mat{\phantom{-}\Phi_2^*(x) & \Phi_1(x) \\ -\Phi_1^*(x) & \Phi_2(x)}. \]
In our case, $\Phi_M$ is obtained by simply omitting the isospinor
$(0,1)^{\rm T}$ in (\ref{felddef}).
The two fermions have equal mass $m=g_{\,\rm Y} \, v / \sqrt{2}$,
due to the $SU(2)_L$ gauge and  $SU(2)_R$ custodial symmetry transformations                           
\bea
\Psi_L(x) &\rightarrow& \Lambda_L(x) \,\Psi_L(x)\,,\quad 
\Psi_R(x) \rightarrow \Lambda_R \,\Psi_R(x)\,, \nonumber \\  [0.1cm]
W_\mu(x) &\rightarrow& \Lambda_L(x) \,\left(W_\mu(x) + \partial_\mu\right) \,
                       \Lambda_L^{-1}(x)\,, \nonumber \\  [0.1cm]
\Phi_M(x)& \rightarrow& \Lambda_L(x) \,\Phi_M(x) \,\Lambda_R^{-1}\,, 
\label{gaugetransf}
\eea
with $\Lambda_L(x), \,\Lambda_R \in SU(2)$.
As far as the fermions are concerned,
the Minkowski metric $\eta^{\mu\nu}$ is taken to have signature $(+---)$
and $\{\gamma^\mu,\gamma^\nu\} = 2\, \eta^{\mu\nu}$.                       
The Dirac matrices $\gamma^0$ and
$\gamma_5 \equiv i \gamma^0 \gamma^1 \gamma^2 \gamma^3$
are taken to be hermitian and the $\gamma^m$ antihermitian.

The field equation derived from this Lagrangian 
is
\beq
i \, \partial_t \Psi = H \, \Psi,
\label{dirac}
\eeq
with the hermitian Dirac Hamiltonian 
\bea
H &=&
-\, i \, \gamma^0\left(
\gamma^\rho\, \partial_\rho+ \rho^{-1} \, \gamma^\phi \, \partial_\phi 
+ ( \gamma^\rho \, W_\rho + \gamma^\phi \, W_\phi ) P_L + \gamma^3\, \partial_z \right)
\nonumber\\[0.1cm]
&& + \, g_{\,\rm Y}\,\gamma^0 \left( \Phi_M P_R + \Phi_M^\dagger P_L \right), 
\label{Hamiltonian}                                              
\eea
for gauge fields $W_0$ and $W_3$ vanishing and 
gamma matrices in cylindrical coordinates
\[
\gamma^\rho \equiv \gamma^1 \cos\phi + \gamma^2\sin\phi\, , \quad 
\gamma^\phi \equiv -\gamma^1\sin\phi + \gamma^2\cos\phi\, .    
\]
The Dirac Hamiltonian (\ref{Hamiltonian}) is real in the sense that\footnote{Quantum 
mechanically, a linear operator $O$ has a complex
conjugate operator $O^{\star}$ defined by $O^{\star}\psi^{\star}$ =
$(O\psi)^{\star}$, for an arbitrary wave function $\psi$.
This operator $O^\star$ is in general different from the adjoint $O^\dagger$,
defined by $(\psi_2, O^\dagger \psi_1)$ = $(O \psi_2, \psi_1)$.
An hermitian operator has $O^\dagger = O$.}
\beq
H^* = Q \, H \, Q^{-1} \; ,
\label{Hreal}
\eeq
with, for example,  in the Majorana representation (all $\gamma^\mu$ imaginary)
\beq
Q = \gamma^0 \, \gamma_5 \, \tau_2 \, .
\label{Q}                                                  
\eeq
Correspondingly, a vector $\Psi$ is called Q-real if $\Psi^* = Q\, \Psi$.
It is not difficult to prove that
non-degenerate eigenvectors of a Q-real hermitian operator (for example, $H$)
can be chosen Q-real, with an unique phase factor up to a sign. This fact
will be important for the Berry phase factor later on.

For static background fields, the stationary solutions of (\ref{dirac}), 
\beq
\Psi(\mathbf{x},t)=\Psi(\mathbf{x})\, \exp(-i\,E\,t) \, ,
\eeq
are given in terms of the solutions to the eigenvalue equation
\beq
H\,\Psi(\mathbf{x})=E\, \Psi(\mathbf{x})\, .
\label{Heigenvalueeq}
\eeq
Remark that the Dirac Hamiltonian (\ref{Hamiltonian}) depends on
the parameters $\mu$ and $\nu$ through the classical background fields
$W$ and $\Phi$ of the $Z$-NCS\p. These background fields can be simplified
significantly if the matrix $\Omega$ in
the definition (\ref{felddef}, \ref{matrizen}) is eliminated by a global
gauge transformation, which does not affect the eigenvalues of $H$. 
As suggested in the previous Section, the resulting background  
may be called the $Z$-flask.

Eigenstates of the transformed Hamiltonian are obtained by using an
ansatz for the fermion fields that takes advantage of the symmetries of
the background fields. The resulting fermion fields are $z$-independent and
possess a continuous rotation symmetry generated by
\beq
K_3\equiv -i\,\partial_\phi + \half \, \Sigma_3 + \half \,\tau_3 \,(P_L-P_R)\, ,
\eeq
together with a certain discrete symmetry.
Here, and in the following, the Pauli matrices  $\tau_a$ act on isospinors, whereas the
$\Sigma_a \equiv \frac{i}{2}\,\epsilon_{abc} \,\gamma_b\,\gamma_c\,$
act on Dirac spinors. The discrete symmetry operator $R_1$ consists, for $\nu=0$,
of a rotation over $\pi$ around the $x^1$-axis and a matching isospin
transformation with $i\tau_1$.  The fermion ansatz and the resulting
differential equations are given in the Appendix A.

\begin{figure}
  \parbox{7.5cm}{
    \rotpicsmall{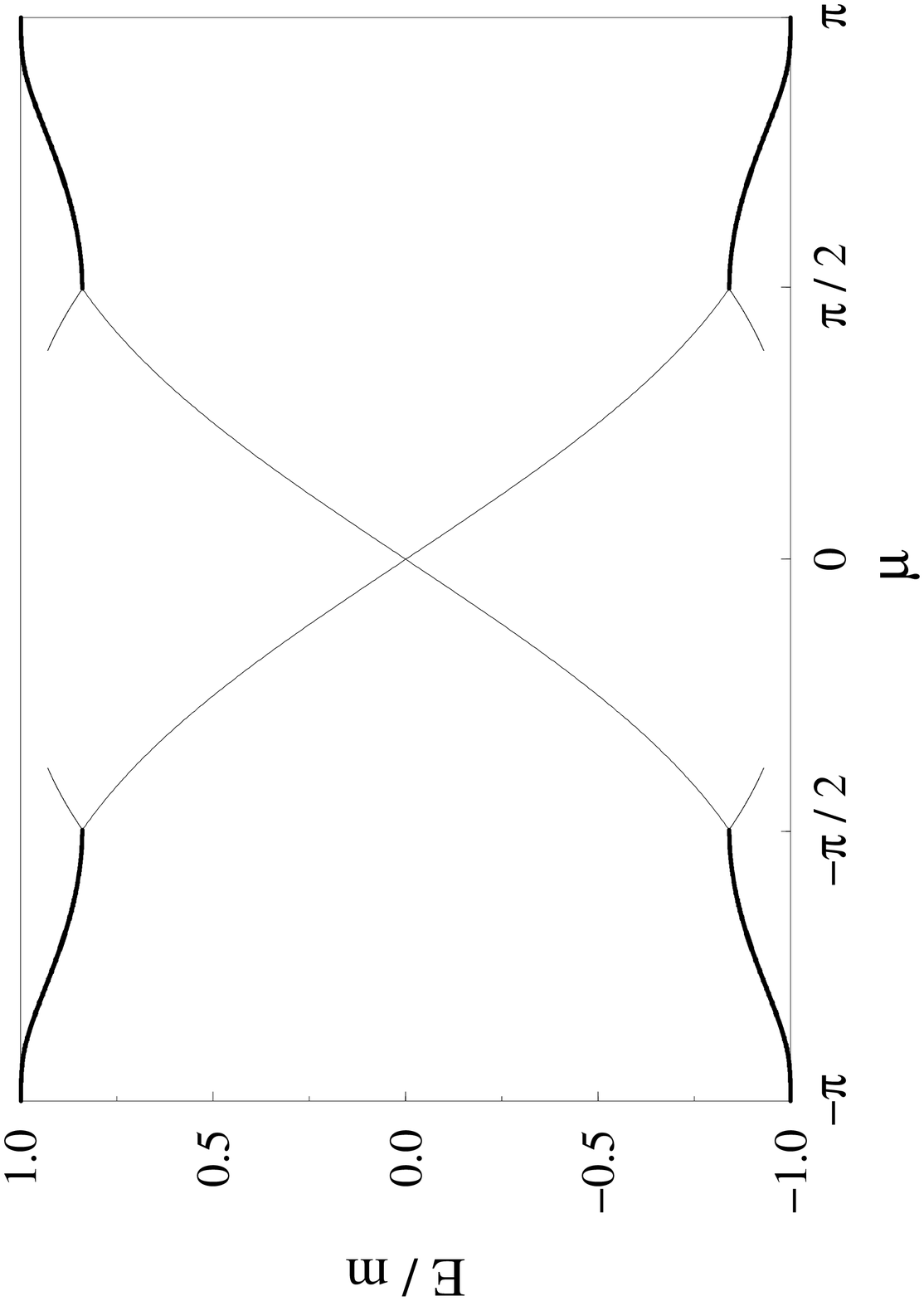}
    \caption{\small
      Energy eigenvalues for $\nu=0$.
      \label{Bildwindeinsemu}
    }
  }
\hfill
  \parbox{7.5cm}{\vspace*{4mm}
    \rotpicsmall{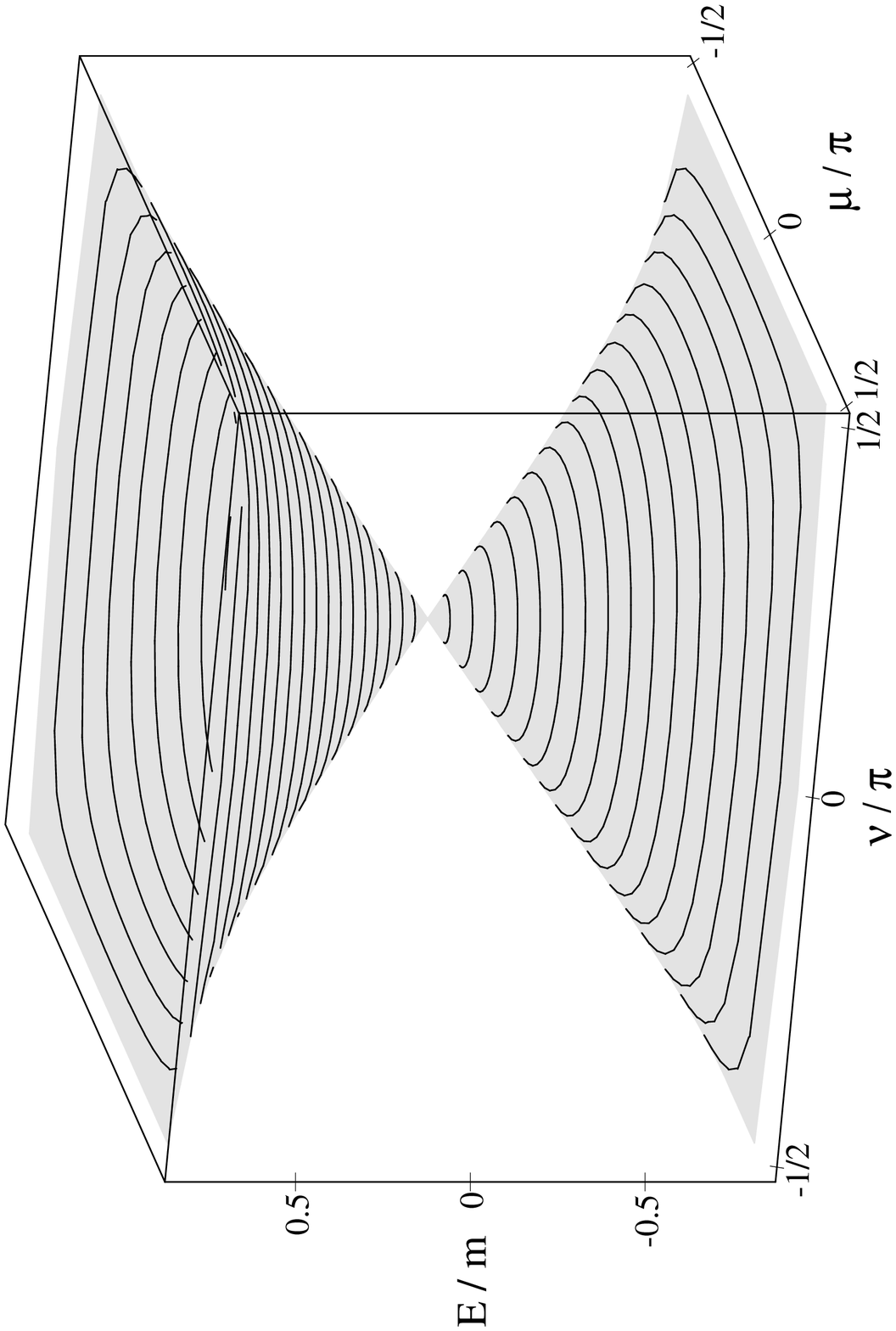}
    \caption{\small
      Energy eigenvalues on the $Z$-NCS\p.
      \label{quadratkegel}
    }
  }
\end{figure}

Setting $\nu=0$ first, the eigenfunctions and eigenvalues of $H$ are obtained
numerically for the case of equal masses  $m=M_H=M_W$.
The phenomenon of spectral flow is observed, with a single pair of eigenvalues
crossing through zero\footnote{These fermion zero modes were already 
discovered in \cite{JR,EP}.}
at $\mu=0$, which corresponds to the $Z$-string
(Fig. \ref{Bildwindeinsemu}). The relevant eigenstates have
$K_3$ eigenvalue $0$ and $R_1$ eigenvalues $\pm 1$. By considering the
continuity of the wave functions, it is shown in Appendix A that the levels 
really cross, instead of being tangent to one other.
The eigenvalues for $|\mu|\ge \pi/2$ are doubly degenerate 
($R_1$ eigenvalues $\pm 1$).                                       

The energy eigenvalues for $\nu\ne 0$ are obtained by taking advantage of
the symmetry of the $Z$-NCS\p ~(or the $Z$-flask, for that matter).
In fact, the Dirac Hamiltonians (\ref{Hamiltonian})
on the special contour $\cos \mu \cos\nu=c$, with $c\in [0,1]$ a constant,
are related by a unitary transformation, so that the eigenvalues are the same. 
This unitary transformation operator $N$ will be given in the next Section.
Plotted against $\mu$ and $\nu$, the eigenvalues display cone-like
spectral flow (Fig. \ref{quadratkegel}) and are everywhere non-degenerate,
except for $\mu=\nu=0$ and the rim of the $\mu,\nu$ square shown.
By continuity, the $Z$-NCS and the $Z$-NCD have essentially the
same cone-like spectral flow.

\section{Berry phase on the $Z$-NCS\p ~and $Z$-NCS}      

We now ask what happens to a real eigenvector of the Dirac Hamiltonian 
$H$ under adiabatic transport along a loop encircling the $Z$-string.
The transported state will differ from the
original state by a phase factor, which consists of the usual dynamical phase
and, potentially, the so-called Berry phase.
Let the loop be parametrized by $\alpha \in [0,2\pi]$ and choose
differentiably for each $\alpha$ a normalized eigenstate\footnote{
For clarity, bra-ket notation is used in this Section.}
$|\Theta(\alpha)\ra$ of
$H(\alpha)$ with non-degenerate eigenvalue $E(\alpha)$.
The implicit condition $|\Theta(0)\ra = |\Theta(2\pi)\ra$ may require 
complex phases in the definition of the eigenstates $|\Theta(\alpha)\ra $,
see below. If $|\Psi\ra_{\mbox{\tiny initial}}= |\Theta(0)\ra $ is transported
along this loop adiabatically (setting $\alpha=2\pi t/T$ 
and taking the total time $T$ large), the result is 
\beq
|\Psi\ra_{\mbox{\tiny final}} = e^{-i\int_0^T {\rm d} t \, E(t)} \,\,
                                e^{i\gamma} \, |\Psi\ra_{\mbox{\tiny initial}}\, , 
\label{Psifinal}
\eeq
with the Berry phase \cite{B}
\beq
 \gamma = i \int_0^{2\pi} {\rm d} \alpha \;
  \la \Theta(\alpha) |\, \pdq{}{\alpha} \,| \Theta(\alpha)\ra\, .
\label{Berryphase}
\eeq

The Berry phase can be calculated directly for the special loops on the
$Z$-NCS\p ~or $Z$-NCS given by $\cos\mu\cos\nu=c$, with $0<c<1$.
In this case one has for the Dirac Hamiltonian
\bea
H(\mu_0,\alpha) &=& N(\alpha)\, H(\mu_0,0)\, N(\alpha)^{-1}\, ,
\label{Hmu0t}
\eea
with 
\beq
N(\alpha) = \exp\left( \, i\,\frac{\alpha}{2} \,\tau_3\,\right)\, ,
\label{N}
\eeq
where $\alpha\equiv 2\pi t/T \in [0,2\pi]$ parametrizes the curve $\cos\mu\cos\nu=c$ and
$H(\mu_0,0)$ is the Dirac Hamiltonian (\ref{Hamiltonian})
corresponding to the point ($\mu=\mu_0\equiv \arccos c$, $\nu=0$) on the curve.
As mentioned before, the eigenvalues of $H$ are constant along these curves
(Fig. \ref{quadratkegel}).
The general solution of the Schr\"{o}dinger-like equation
(\ref{dirac}, \ref{Hmu0t})  is
\beq
|\Psi(\mu_0,t)\ra =
       e^{-i\,\tilde{E}(\mu_0)\,t} \, N(2\pi t/T) \, |\Psi(\mu_0,0)\ra \, ,
\label{Psisolution}
\eeq
with
\beq
\left[ H(\mu_0,0)+ (\pi/T)\,\tau_3\, \right] \, |\Psi(\mu_0,0)\ra =
\tilde{E}(\mu_0) \, |\Psi(\mu_0,0)\ra \, .
\label{Ebareigenvectors}
\eeq
The normalized eigenvectors (\ref{Ebareigenvectors}) approach
in the adiabatic limit ($T \rightarrow \infty$) those of the
initial Hamiltonian
\beq
H(\mu_0,0) \, |\Psi(\mu_0,0)\ra = E(\mu_0) \, |\Psi(\mu_0,0)\ra \, ,
\eeq
so that
\beq
\tilde{E}(\mu_0) =
     E(\mu_0) + (\pi/T) \: \la \Psi(\mu_0,0)|\,\tau_3\,|\Psi(\mu_0,0)\ra \, .
\label{Etilde}
\eeq
The $\tau_3$ expectation value in (\ref{Etilde}) vanishes identically, because 
the state $|\Psi(\mu_0,0)\ra$ considered has $R_1$ eigenvalues $\pm 1$  and the
anticommutator $\{R_1,\tau_3\}$ vanishes, see Appendix A for further details.
The solution (\ref{Psisolution}) for $t=T$, with $N(2\pi)$ = $-N(0)$ = $-\id_2 $, 
gives then the Berry phase
\beq
\gamma = \pi
\label{Bphaseminpi}
\eeq
for adiabatic transport along the curve $\cos\mu\cos\nu=c$.

This non-vanishing Berry phase (\ref{Bphaseminpi}) is essentially due to the         
fact that the transformation matrix $N(\alpha)$ of the Hamiltonian
(\ref{Hmu0t}) runs from one element in the center
of the group $SU(2)$ to the other.
Of course, this is only possible for fermions in $SU(2)$ representations of
half-integer isospin ($I=1/2$ for the doublet here).
Note, finally, that the Berry formula (\ref{Berryphase}), with
the differentiable (complex) choice of eigenstates
\beq
 | \Theta(\alpha)\ra = e^{-i\,\alpha /2} \, N(\alpha) \, |\Psi(\mu_0,0)\ra \, ,
\eeq
gives the same result $\gamma=\pi$.
The advantage of the general solution (\ref{Psisolution}) is that it also
applies to the case of degenerate eigenvalues $E(\alpha)$.

The result (\ref{Bphaseminpi}) for the Berry phase over these special loops on
the \ncs ~can be extended to arbitrary loops by the following argument.
As mentioned in Section 3, the Dirac Hamiltonian obeys the reality condition
(\ref{Hreal}). 
Moreover, the eigenvalues under consideration are non-degenerate, see
Fig. \ref{quadratkegel}. Under these
conditions, the Berry phase factor $e^{i\gamma}$ is necessarily $\pm 1$,
as follows from the remarks below (\ref{Q}). 
Since these allowed values are isolated points in $\R$, the Berry phase factor
(which is a continuous functional of the loop) cannot change if the loop is
continuously deformed. Thus, every loop on the $Z$-NCS (or $Z$-NCS\p)
that winds around the $Z$-string exactly once,   
and does not touch the degeneracy points ($\mu=\nu=0$ or $[\mu\nu] = \pi/2$),
leads to the same Berry phase factor $-1$.
More generally, the Berry phase $\gamma_{\,\rm C}$ for any closed curve C
on the $Z$-NCS with the degeneracy points omitted, which corresponds to a
cylinder topologically, is given by
\beq
e^{\, i\, \gamma_{\,\rm C}} = \, (-1)^{\,n_{\,\rm C}} \, ,            
\label{gammaC}
\eeq
with $n_{\,\rm C}$ the winding number of the curve C on the cylinder.

The true origin of the Berry phase factor $-1$ is
the fermion degeneracy in the $Z$-string background.
This is clarified in Appendix B by a different derivation of the Berry phase,
in terms of a solid angle in parameter space.

\section{Berry phase on the $Z$-NCD}

In the previous Section we have established a non-vanishing Berry phase $\gamma=\pi$
for loops on the $Z$-NCS\p ~or $Z$-NCS (Fig. \ref{skizze}ab)
that circumnavigate the $Z$-string once. For a loop close
to the degeneracy point (i. e. the $Z$-string)
we have calculated, in Appendix B, the Berry phase more generally, in terms of an
abstract solid angle in parameter space. The Berry phase factor $e^{i\gamma}=-1$
for this small loop on the $Z$-NCS carries over to a small loop on the $Z$-NCD 
(Fig. \ref{skizze}c).  The reason is that, on the one hand,
the background fields can be mapped into each other continuously, and, on the other hand,
the Dirac Hamiltonian is essentially real, so that the Berry phase factor $e^{i\gamma}$
stays discrete ($\pm 1$).
See also the discussion in the paragraph leading up to (\ref{gammaC}).
The implicit assumption here is that there are no further
degeneracies, which is certainly the case for loops close enough to the $Z$-string.
 
\begin{figure}
  \centerline{
    \mbox{
      \parbox[t]{7cm}{
        \unitlength1cm
        
        \begin{picture}(7,5)
          \put(-3,7.5){\includegraphics{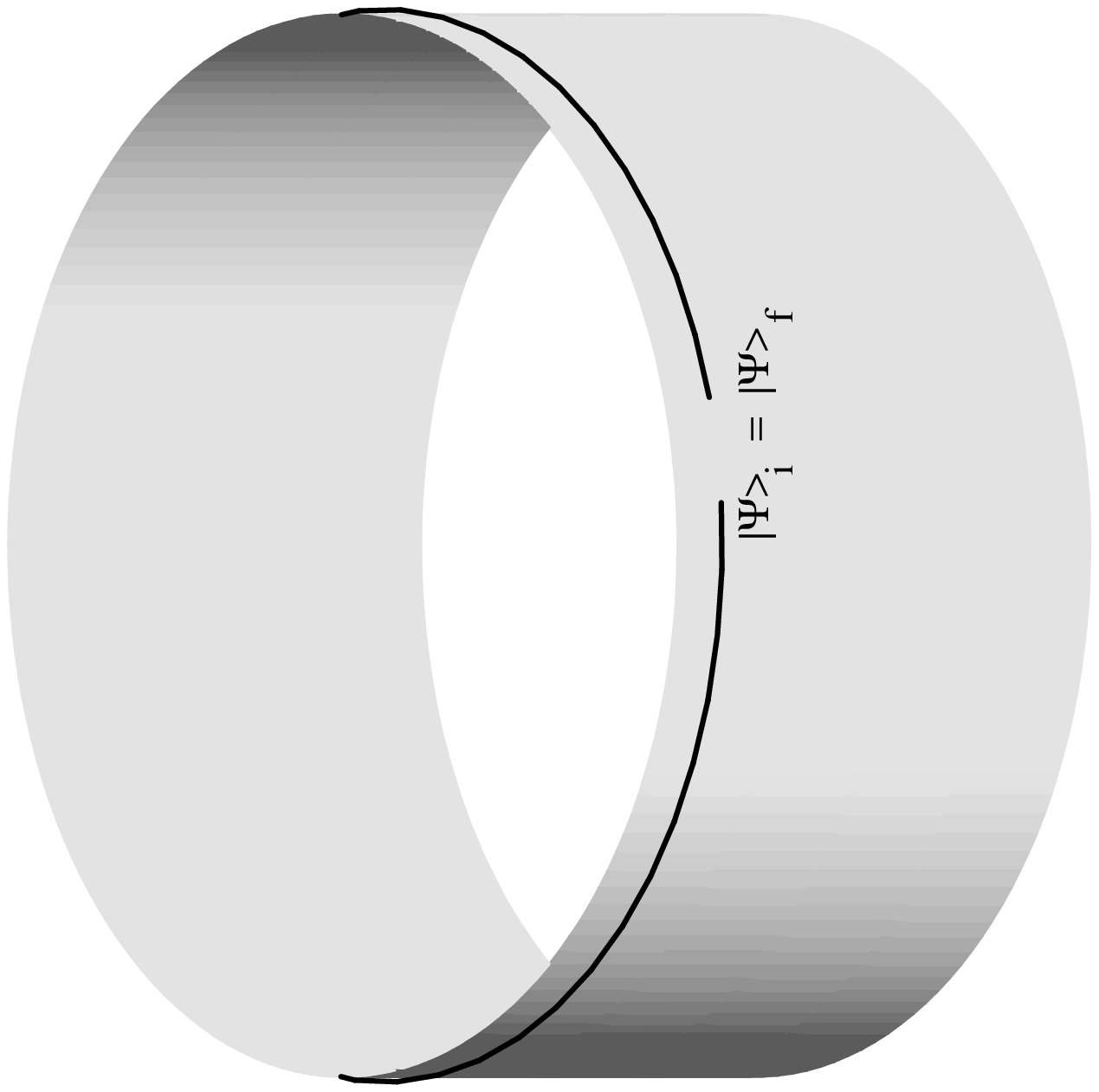}}
        \end{picture}\\
        }
      }
    \parbox[t]{7cm}{
      \unitlength1cm
      
      \begin{picture}(7,5)
        \put(-3,7.5){\includegraphics{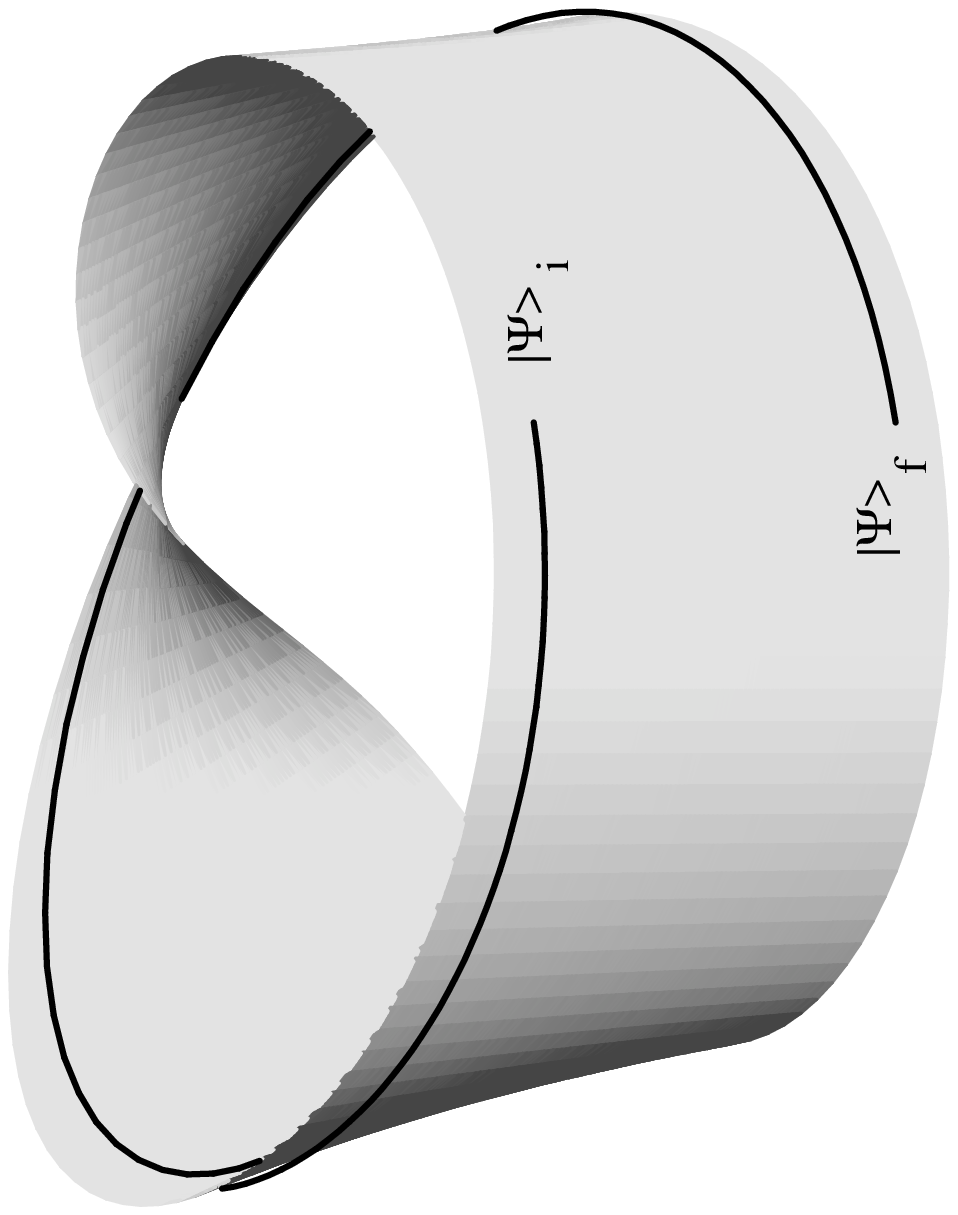}}
      \end{picture}\\
      \centerline{}
      }
    }
\caption{Cylinder and M{\"o}bius bundle over different gauge orbits,
         with the line representing a
         normalized real eigenstate of the Dirac Hamiltonian.}
\label{moebiusbild}
\end{figure}

There is thus a Berry phase factor $e^{i\gamma}=-1$ for a small loop around the top of the $Z$-NCD.
This small loop can be pulled towards the vacuum continuously ($\eta \rightarrow \pi/2$),
with the discrete Berry phase factor remaining at the value $-1$.
In this way we end up with a \ncl ~of vacuum bosonic field configurations 
(\ref{ZNCDfields}, \ref{ZNCDmatrices}), for $\eta=\pi/2$ and $\omega \in [0,2\pi]$, 
which gives a non-trivial Berry phase factor $-1$.
The real sub-bundle over this particular gauge orbit is {\em twisted\/},
so that instead of a trivial cylinder bundle there is, in fact,
a M\"{o}bius bundle (Fig. \ref{moebiusbild}).
This non-trivial Berry phase factor of first-quantized fermion states over the
vacuum gauge orbit is a necessary, but not sufficient, condition for the
global anomaly of the purely left-handed gauge field theory to be
discussed in the next Section.
Note that for the numerical calculations of Section 3 the right-handed     
fermions were introduced in order to have a mass gap, but for the
Berry phase calculation in this and the previous Section they can
be left out from the start.\footnote{With massless fermions it may be
necessary to have also a finite range for the $x^1$ and $x^2$ coordinates.}
For completeness, we discuss in Appendix C the Berry phase
on the \ncs ~with winding number $n=2$, which leads to a gauge orbit
with trivial bundle.

At this point, we like to mention two                           
issues that deserve further study. The first issue is wether or not
there is an index theorem responsible for the observed spectral flow
of Dirac eigenvalues over the \ncs ~(Fig. \ref{quadratkegel}).
For the Witten anomaly \cite{W} to be discussed in the following Section
there is the so-called mod 2 Atiyah-Singer index
theorem and in our case the standard Atiyah-Singer index
theorem \cite{C,JRe} may be expected to play a role.
In fact, the matrix $M$ as given in (\ref{matrizen}) has the same structure
as the corresponding matrix of the BPST instanton \cite{BPST}.
The second issue is the removal of the range limitation on the $x^3$
coordinate ($\ell_3 \rightarrow \infty$). 
A related point is the choice of boundary conditions on the fields
at $x_3$ $=$ $\pm \, \ell_3 /2$ (implicitely we have taken free boundary conditions). 
All the results presented in this paper are formally independent of $\ell_3$. In particular,
the reality condition (\ref{Hreal}) on the Dirac Hamiltonian holds generally.
Still, the existence and the nature of the $\ell_3$ limit need to be established
rigorously.  Pending this proof, we keep the range $\ell_3$ large, but finite.
 
\section{Global gauge anomaly}

One speaks of a gauge anomaly if there occurs, upon quantization of a chiral
gauge field theory, an insurmountable obstacle to maintaining gauge invariance. 
Concretely, the quantization in the hamiltonian formulation 
may proceed in two steps. First, the fermions are quantized (``filling the Dirac
sea'') in the background of the classical boson fields. 
Second, the boson fields themselves are quantized. 
What may then happen is the following (quoting liberally from
\cite{NAG}): Quantizing the fermionic
matter in the presence of background gauge fields, the resulting
family of quantum theories in general realizes its classical gauge
symmetry via a perfectly good ray representation. 
As far as the fermions are concerned there is nothing wrong with gauge
symmetry. If, however, the phases in the ray representation are
topologically unremovable, then they prevent us from implementing the symmetry
in the fully quantized theory, and in particular from
imposing the constraint of gauge invariance (Gauss' law) on the
physical quantum states. 
We will now show that this is precisely what follows from the results of the previous Sections. 

The central object for the anomaly discussion here is the gauge-covariant
left-handed Dirac operator                 
\beq
\not\!\!D_L  \equiv \gamma^\mu \, (\partial_\mu + W_\mu)\, P_L \, ,
\label{DslashL}
\eeq
with $W_\mu$ antihermitian and in the Lie algebra of the 
gauge group.\footnote{The anomaly in the euclidean path integral formulation comes from the 
ambiguous determinant (product of eigenvalues) of $i\!\!\not\!\!\!D_L$.
The operator $i\!\!\not\!\!\!D_L$ has, strictly speaking, no eigenvectors,
as it maps left-handed fields to right-handed fields and {\em vice versa\/}.}
Consider then $SU(2)$ \YMHth ~with a single doublet of left-handed fermions
(and no right-handed fermions) on a flat $3+1$ dimensional space-time manifold with
a large, but finite, range $\ell_3$ for the the spatial 
$x^3$ coordinate.  In the previous Section we discovered that for a particular loop 
of gauge transformations of the classical bosonic vacuum fields the
first-quantized fermion states acquired a
topological phase factor, namely the Berry phase factor $e^{i\gamma} = -1$.
More importantly, there is a \emph{single} pair of fermionic
levels crossing at $E=0$ in the $Z$-string background
(see Fig. \ref{Bildwindeinsemu} and Appendix A).
For the second-quantized vacuum state this results in a
M\"{o}bius bundle structure over the gauge orbit, which interferes
with the definition of physical states, see Appendix D.
Therefore, $SU(2)$ \YMH ~quantum field theory  with a single doublet of
massless left-handed fermions (Weyl spinors) is anomalous.
Without exact gauge invariance, the theory is most likely inconsistent.

Concretely, our loop of gauge transformations, given by
(\ref{ZNCDfields}, \ref{ZNCDmatrices})  for $\eta=\pi/2$ and loopparameter
$\omega \in [0,2\pi]$, is independent
of the spatial $x^3$ coordinate and the topologically non-trivial map is
\beq
  S_1 \times S_2 \sim S_3 \longrightarrow SU(2) \sim S_3 \, ,
  \label{Uvac}
\eeq
with winding number $n=1$.\footnote{
For the case of winding number $n=2$ there are two pairs of fermionic
levels crossing at $E=0$ (see Appendix C) and the second-quantized vacuum bundle is trivial.
In this Section only winding number $n=1$ is considered (later on $n$ will
denote the dimension of the $SU(2)$ representation of the fermions).}
This may be compared to the Witten global $SU(2)$ anomaly in the
hamiltonian formulation \cite{W}, where the loop of gauge transformations
depends on all three spatial coordinates and the non-trivial map is
\beq
  S_1\times S_3 \sim S_4 \longrightarrow SU(2) \sim S_3 \, ,
  \label{UWitten}
\eeq
which is the suspension of the Hopf map $S_3 \rightarrow S_2$.
Both maps (for the same gauge group $G=SU(2)$) can indeed be non-contractible,
since the respective homotopy groups are non-trivial
$\pi_3(SU(2))= \Z$ and $\pi_4(SU(2))= \Z_2$,
where $\Z$ denotes the group of integers and $\Z_2$ the integers modulo $2$.
Moreover, both \ncls ~pick up a Berry phase factor $-1$, because they
encircle a degeneracy point in \cs ~corresponding to 
the $Z$-string \cite{N} and the \Sstar ~sphaleron \cite{K}, respectively. 
Thus, chiral SU(2) \YMH ~quantum field theory with 
an odd number of massless left-handed fermion doublets is ruled out on both counts.

Up till now we have considered chiral $SU(2)$ \YMHth, but our results may also
apply to theories without Higgs. It is clear that for massless fermions
the Berry phase depends on the gauge fields only.
As long as there remains an encircled  degeneracy point
(most likely, guaranteed by an index theorem), the bundle over
the loop of vacuum gauge fields stays twisted, even in the
absence of the Higgs fields. Hence, pure $SU(2)$ gauge field theory with
an odd number of massless left-handed fermion doublets is also expected to 
suffer from the new global anomaly (in addition to the Witten anomaly,
of course).

We can make two further generalizations. First, consider for the 
background fields of the \ncs ~(\ref{ZNCSfields}) massless
left-handed fermions  in $SU(2)$ representations larger than the
doublet representation (isospin $I=1/2$).
A straightforward, but tedious, calculation gives for a single irreducible representation of isospin $I$ 
and dimension $n = 2\,I +1$ the following number of pairs of fermionic levels crossing at $E=0$ :
\beq
N_{\rm pair}     = \left\{
\begin{array}{ll}
n^2 /\, 4        & \quad \mbox{if $\; n=2\,l       $}    \\[0.1cm]
(n^2 - 1)\,/\, 4 & \quad \mbox{if $\; n=2\,l+1 \, ,$}
\end{array}
                   \right.
\label{Npairs}
\eeq
with $l$ a positive integer. There is then an odd number of pairs crossing
for isospin values
\beq
I= \frac{4\,k+1}{2} \, , \quad k=0,1,2, \ldots \; .
\label{isospinors}
\eeq
Therefore, $SU(2)$ \YMparH ~quantum field theory with massless left-handed
fermions  in a single  irreducible representation  of isospin (\ref{isospinors}) 
has  a twisted vacuum bundle and the theory suffers from the new
global anomaly.\footnote{It is important to
verify this result by other methods, for example euclidean path integrals 
(paying attention to the relevant fermion zero modes).}
Second, consider gauge groups $G$ other than $SU(2)$ and massless left-handed
fermions in corresponding representations.  
For \emph{any} compact connected simple Lie group $G$ 
the third homotopy group is again non-trivial
\beq
  \pi_3(G) = \Z \, .
  \label{pi3G}
\eeq
Furthermore, it is known that any continuous mapping
of $S_3$ into $G$ can be continuously deformed into a mapping into an
$SU(2)$ subgroup \cite{C}.
Hence, all constructions in $SU(2)$ \YMth ~carry over to
theories with gauge group $G\supset SU(2)$ and the same twist factor
$-1$ appears for each embedded fermion doublet (or other anomalous isospinor).
If the gauge group $G$ acts on left-handed fermion fields only, 
a further condition for the new global anomaly is then that the \emph{total}
number $N_{I=2k + 1/2}$ of irreducible representations (\ref{isospinors})
of the $SU(2)$ subgroup considered be odd
\beq
  (-1)^{\,N_{I=2k + 1/2}} = -1 \, .
  \label{Nodd}
\eeq
Chiral gauge field theories fulfilling the conditions (\ref{pi3G}, \ref{Nodd})
suffer thus from the same global anomaly (twisted bundle) 
due to the $Z$-string-like fermion degeneracy as the doublet $SU(2)$ \YMHth. 

Remarkably, the same isospinors (\ref{isospinors}) are singled out by the Witten
anomaly.
In fact, for $SU(2)$ generators $T_a$ with algebra
$[\,T_a\, , \,T_b\,]$ $=$ $\epsilon_{abc}\, T_c$ the Witten anomaly \cite{W} requires
$N_0$ $\equiv$ $- 2 \,{\rm Tr} \, (T_3)^2$ $=$ $n\,(n^2-1)/6$ to be an odd integer,
corresponding to representations of dimension $n$ $=$ $4\, k + 2$ and
isospin $I=(4\,k+1)/2$ for $k=0,1,2, \ldots\;$.\footnote{
$N_0$ turns out to be equal to the number of
fermion zero modes \cite{JRe} in the 4-dimensional background of the BPST
instanton \cite{BPST}. Physically, this
connection is made plausible by Goldstone's derivation \cite{J} of
the Witten $SU(2)$ anomaly. Note that $N_0(n)$ is in general different from
$N_{\rm pair}(n)$ as given by (\ref{Npairs}),
even though both find the same isospin values (\ref{isospinors}) through
$(-1)^{N_0}$ $=$ $(-1)^{N_{\rm pair}}$ $=$ $-1$. }
However, the new global anomaly does rule out theories allowed in principle
by Witten's global anomaly, because $\pi_4(G)$ is non-trivial only in certain cases,
whereas (\ref{pi3G}) holds generally.
A simple example is $SU(3)$ \YMparHth\ with an odd number of massless left-handed
fermion triplets. This theory has no genuine Witten anomaly
(the fourth homotopy group of $SU(3)$ is trivial),
but does satisfy the conditions (\ref{pi3G}, \ref{Nodd}) for
the new global anomaly.  Of course, chiral $SU(3)$ gauge theory
with \emph{any} number of left-handed fermion triplets also has the \emph{local}
(perturbative) Bardeen anomaly \cite{Bar}.
This brings us to the following question: which chiral \YMths ~are safe from
perturbative Bardeen anomalies
(either intrinsically or by
cancellations between the different representations), but do suffer from the
new global anomaly? The group-theoretic results of \cite{O}
appear to rule out all compact simple gauge groups
$G$ with $\pi_4(G) =0$. This leaves essentially the $Sp\,(N)$ groups
($Sp\,(1)$ $=$ $SU(2)$), which are free from perturbative anomalies,
but can have the new global anomaly (and the Witten anomaly)
provided condition (\ref{Nodd}) holds.

To summarize, the main idea of this paper is to consider in $3+1$ dimensional 
chiral Yang-Mills(-Higgs) quantum field theory in the hamiltonian
formulation a non-contractible loop of gauge transformations with 
reduced dependence on the spatial coordinates.
A non-trivial Berry phase factor $-1$ can only occur
if the loop encircles a configuration with degenerate fermions
(in this paper, the embedded $Z$-string).
Depending on the fermion representations present, this may then result in
a twisted bundle over the gauge orbit, which signals the presence of a 
new kind of global gauge anomaly in the theory.
\vspace{1\baselineskip}

One of us (FRK) acknowledges stimulating discussions with the participants
of the $9^{\rm th}$ Workshop on Physics Beyond the Standard Model,
Bad Honnef, March 3--6, 1997.

\begin{appendix}

\section{Fermion ansatz and numerical solution}
\renewcommand{\theequation}{A.\arabic{equation}}
\setcounter{equation}{0}

\renewcommand{\thefigure}{A.\arabic{figure}}
\setcounter{figure}{0}

In this Appendix we give the details of the ansatz for isodoublet fermions
in the $Z$-NCS\p ~background  and present the numerical solution. 
Also, we show that the energy levels
really cross at the top of the \ncs. The relevance of this will be explained at the end.

The classical background fields considered here are given by the
$\nu=0$ slice of $Z$-NCS\p ~(\ref{felddef}, \ref{matrizen}),
with the matrix $\Omega$ eliminated by a global gauge transformation.
Furthermore, we take in this Appendix the 2-dimensional version of
the Dirac Hamiltonian (\ref{Hamiltonian}), that is without the term
$-i\,\gamma^0 \,\gamma^3 \,\partial_z$.                        
The resulting Dirac Hamiltonian commutes with
\beq
K_3 \equiv L_3 +  \half \,\Sigma_3 +\half \,\tau_3 \,(P_L-P_R)
\eeq
and
\beq
R_1 \equiv \exp\,       [-i\pi( L_1+\frac{1}{2}\,\Sigma_1+\frac{1}{2}\,\tau_1)]\,,
\eeq
where $L_k$ and $\Sigma_k$ are the standard orbital and spin angular momentum
operators $L_k$ $\equiv$ $-i \, \epsilon_{klm}\, x_l \, \partial_m$ and
$\Sigma_k$ $ \equiv$ $ \frac{i}{2}\,\epsilon_{klm} \,\gamma_l\,\gamma_m\,$.
The action of $R_1$ on the 2-dimensional coordinates is simply to change
the azimuthal angle $\phi$ into $-\phi$.
Since $K_3$ and $R_1$ do not commute, it is in general
not possible to find common eigenfunctions.  However, common
eigenfunctions do exist in the subspace of vanishing $K_3$ eigenvalue,
since $\{K_3,R_1\}=0$. These eigenfunctions are
 \begin{eqnarray}
   \Psi_1(\rho, \phi) &=& i \,G_L \,( |L\ua d\ra - |L\da u\ra )
                          +   F_L \,( e^{i\phi} |L\da d\ra - e^{-i\phi} |L\ua
                          u\ra ) + \nonumber \\ [0.1cm]
                      & &   G_R \,( |R\da d\ra + |R\ua u\ra )
                          + i \, F_R \,(e^{-i\phi} |R\ua d\ra + e^{i\phi} |R
                          \da u\ra )\, , \nonumber\\ [0.2cm]
   \Psi_2(\rho, \phi) &=& i \, G_L \,( |L\ua d\ra + |L\da u\ra )
                          - F_L \,   ( e^{i\phi} |L\da d\ra + e^{-i\phi} |L\ua
                          u\ra ) + \nonumber\\ [0.1cm]
                      & &   G_R \,( |R\da d\ra - |R\ua u\ra )
                          - i \,F_R \,(e^{-i\phi} |R\ua d\ra - e^{i\phi} |R
                          \da u\ra )\, ,  \label{ansatz}
 \end{eqnarray}
where $G_L$, $G_R$, $F_L$ and $F_R$ are real functions of $\rho$ (with suitable boundary
conditions) and the kets stand for constant normalized eigenvectors of $\gamma_5$,
$\Sigma_3$ and $\tau_3$, in an obvious notation.
The eigenfunctions have the following eigenvalues:
\beq
\begin{array}{rclcrcl}
 K_3 \,\Psi_1 &=& 0\,, & \quad & R_1 \,\Psi_1 &=& +\,\Psi_1\,, \\
 K_3 \,\Psi_2 &=& 0\,, & \quad & R_1 \,\Psi_2 &=& -\,\Psi_2\,.
\end{array}
\eeq

If $\Psi_1$ is inserted into the eigenvalue equation (\ref{Heigenvalueeq}) of the
Dirac Hamiltonian, the following equations are obtained:
\begin{eqnarray}
\partial_\rho G_{L}&=&
  -\,\frac{f}{\rho}\cos^2\mu \, G_{L}+\frac{f}{2 \rho} \sin 2\mu \, F_{L}+m\,h \cos\mu
  \, G_{R}+m\,h\sin\mu \, F_{R} +E \, F_{L}\,, \nonumber\\
\partial_\rho F_{L}&=&
  -\,\ri \,F_{L}+\frac{f}{\rho}\cos^2\mu \,F_{L}+\frac{f}{2 \rho}\sin 2\mu
  \,G_{L}+m\,h\cos\mu F_{R}-m\,h\sin\mu \,G_{R} -E \,G_{L}\,, \nonumber\\
\partial_\rho F_{R}&=&
  -\,\ri \,F_{R}+m\,h\cos\mu \,F_{L}+m\,h\sin\mu \,G_{L} +E
  \,G_{R}\,, \nonumber\\
\partial_\rho G_{R}&=&
  -\,m\,h\sin\mu \,F_{L} +  m\,h\cos\mu \,G_{L} -E \,F_{R} \, ,
\label{WindeinsGln}                                
\end{eqnarray}
with $f(\rho)$ and $h(\rho)$ the profile functions \cite{NO,VS}
of the $Z$-string gauge and Higgs fields, respectively,
and $m = g_{\,\rm Y} \,v/\sqrt{2}$ the fermion mass in the Higgs vacuum.     
For $\mu=0$ one recovers the $Z$-string fermion zero mode \cite{JR,EP},
in terms of two functions $G_L$ and $G_R$ ($F_L=F_R=0$).

It is not necessary to investigate the second ansatz $\Psi_2$ seperately.
All solutions with $R_1$ eigenvalue $-1$ can be obtained from
those with $R_1$ eigenvalue $+1$ by applying the operator
$-\Sigma_3 \gamma_5$,
since $-\Sigma_3  \gamma_5$ anticommutes with $R_1$ and $H$.
Furthermore, it it sufficient to consider the case $\mu\ge 0$. If $\mu$ is replaced
by $-\mu$ in (\ref{WindeinsGln}), the change in the equations can be compensated
by changing simultaneously the eigenvalue $E$ to $-E$ and the functions
$F_{L}$, $F_{R}$ to  $-F_{L}$, $-F_{R}$. 
This shows that simple sign changes can turn
a solution for $\mu$, with energy $E$, into a solution for $-\mu$,
with energy $-E$.
 
\begin{figure}
\rotpicsmall{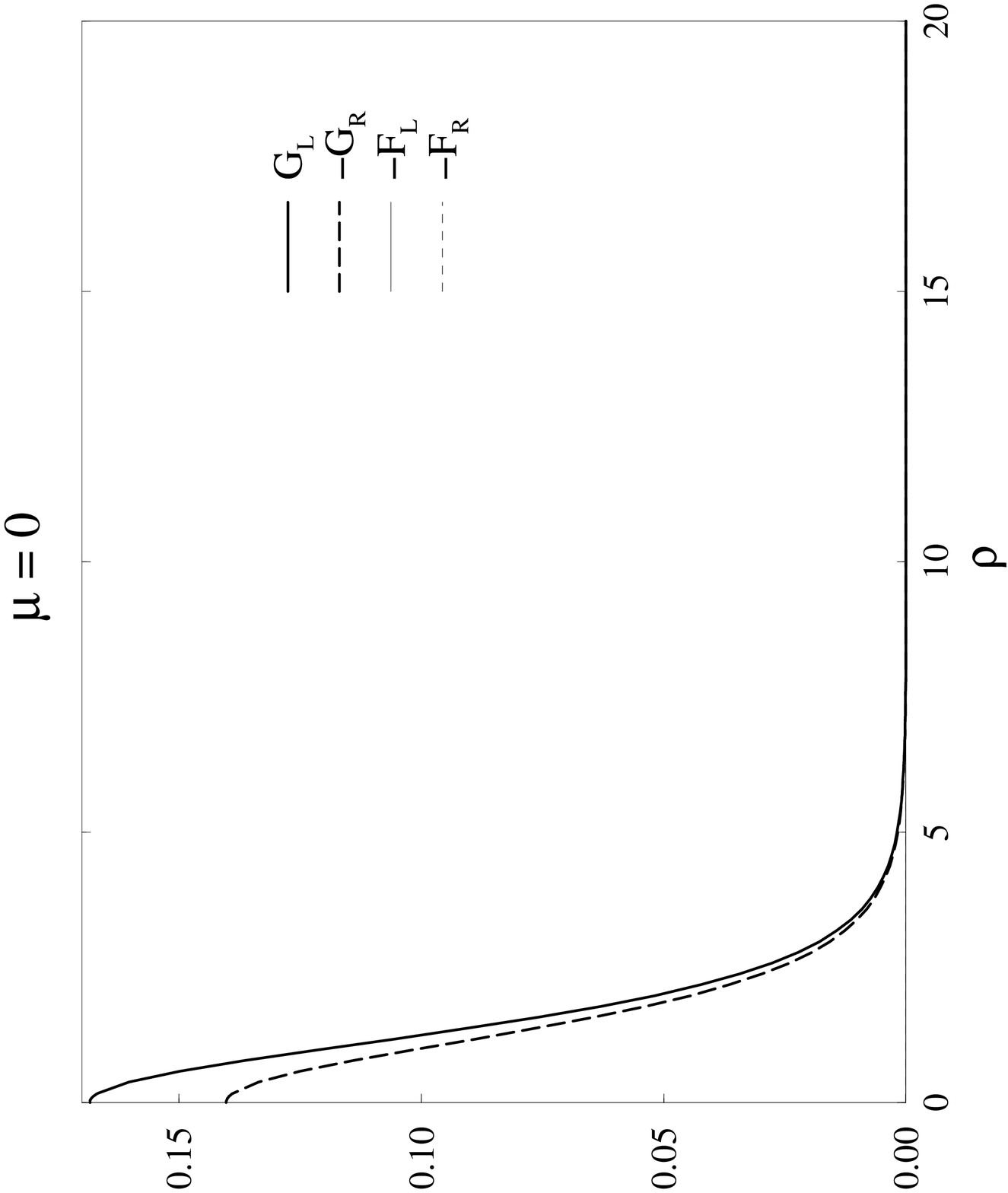}
\rotpicsmall{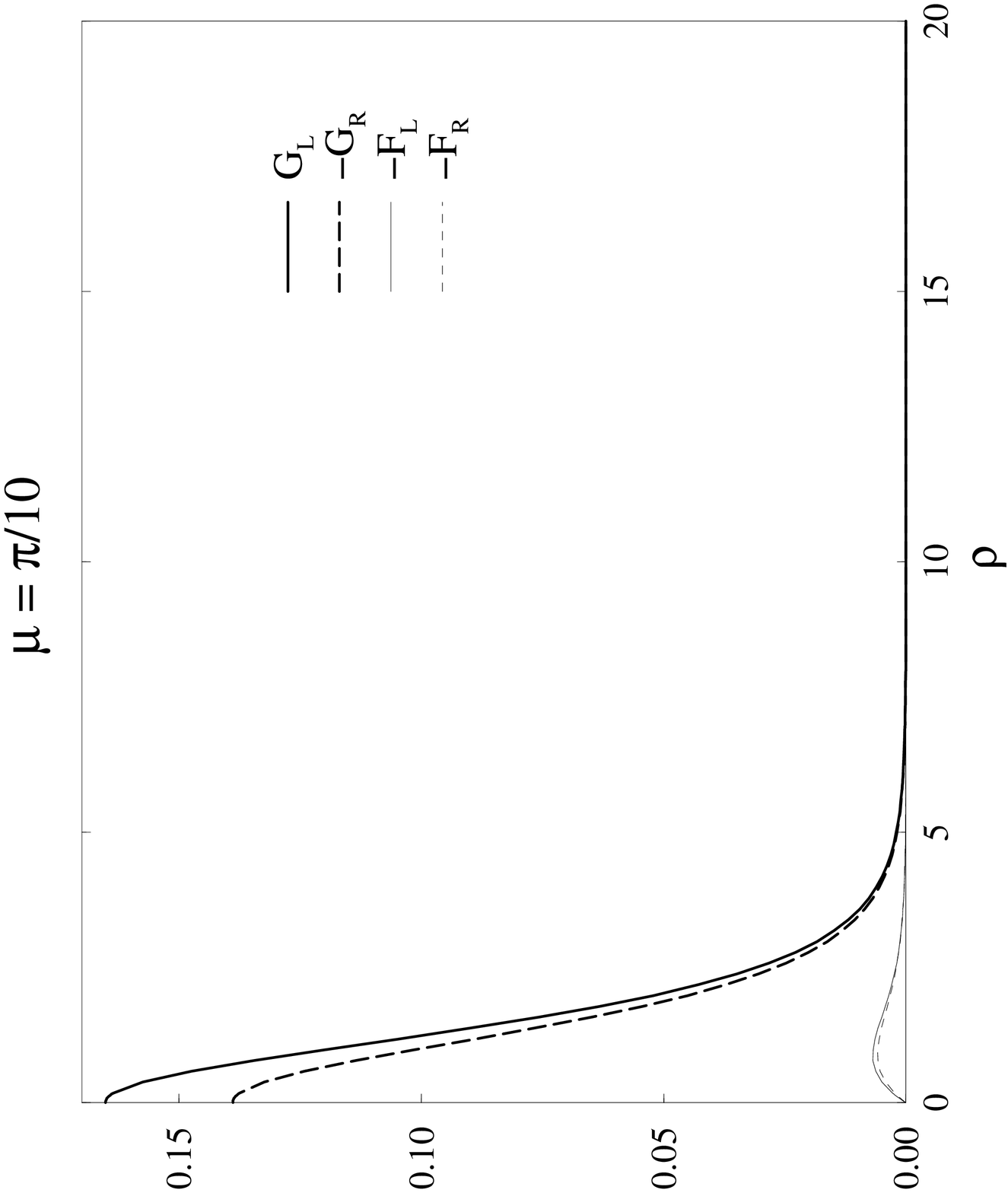}\\[0.2cm]

\rotpicsmall{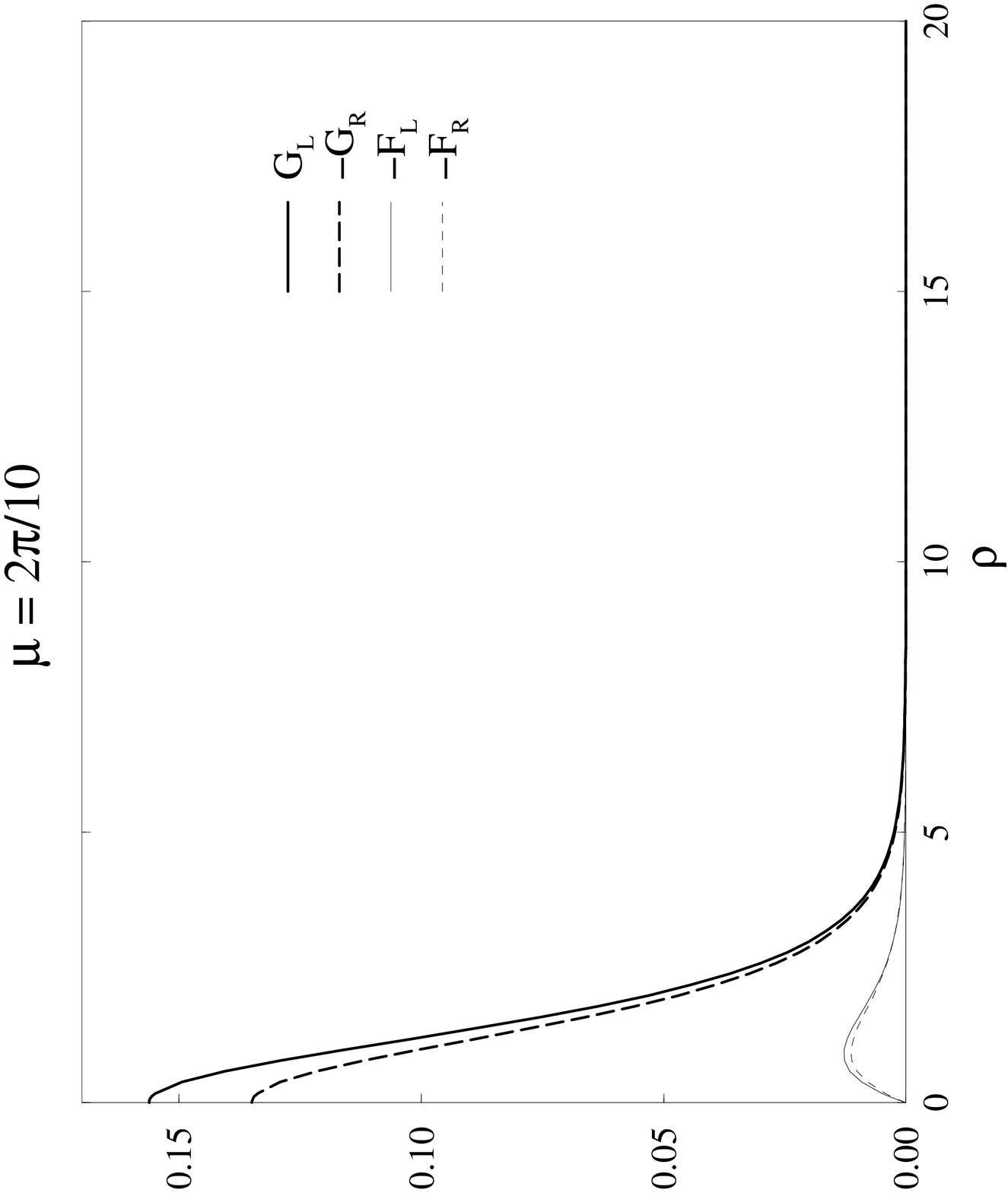}
\rotpicsmall{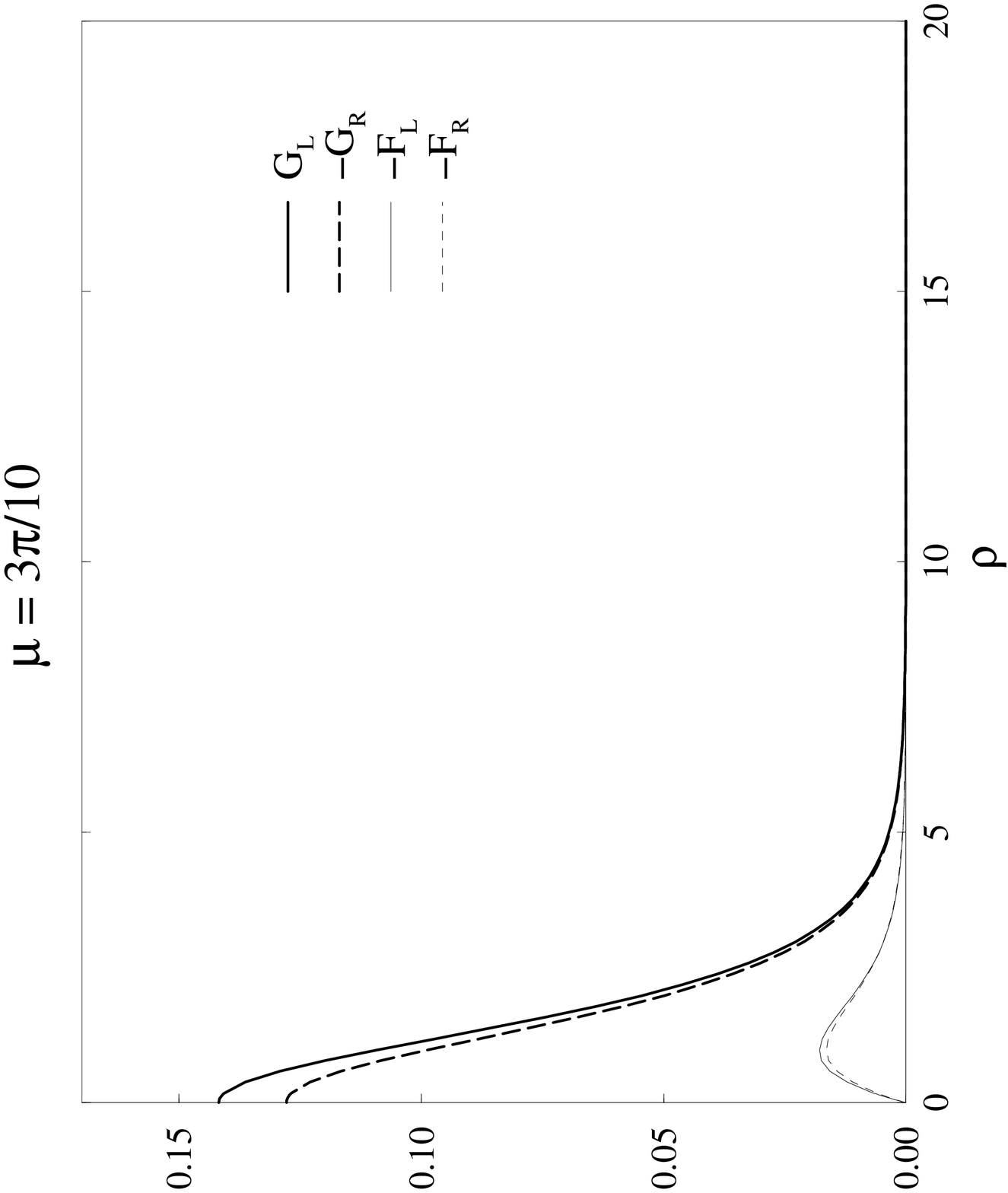}\\[0.2cm]

\rotpicsmall{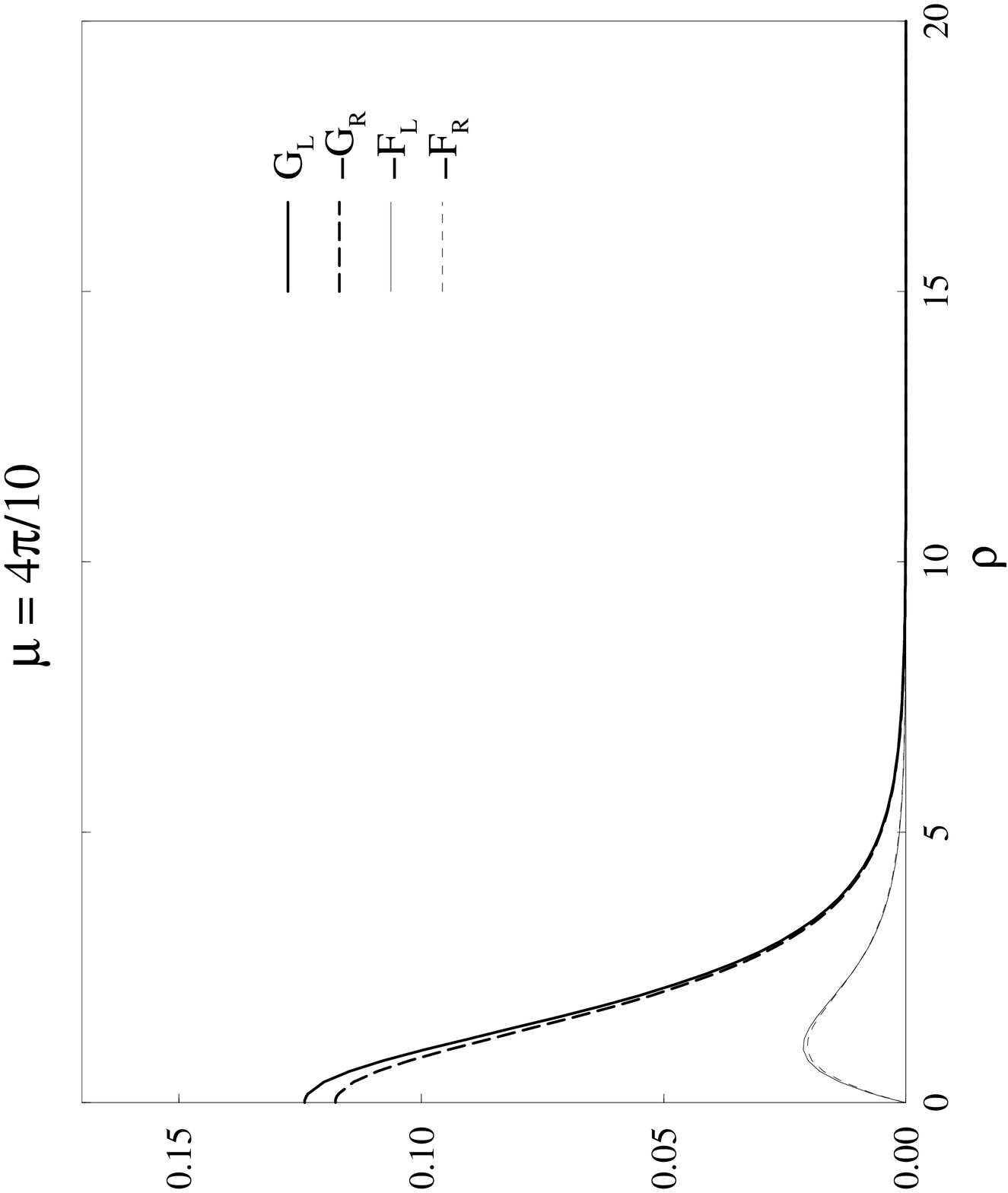}
\rotpicsmall{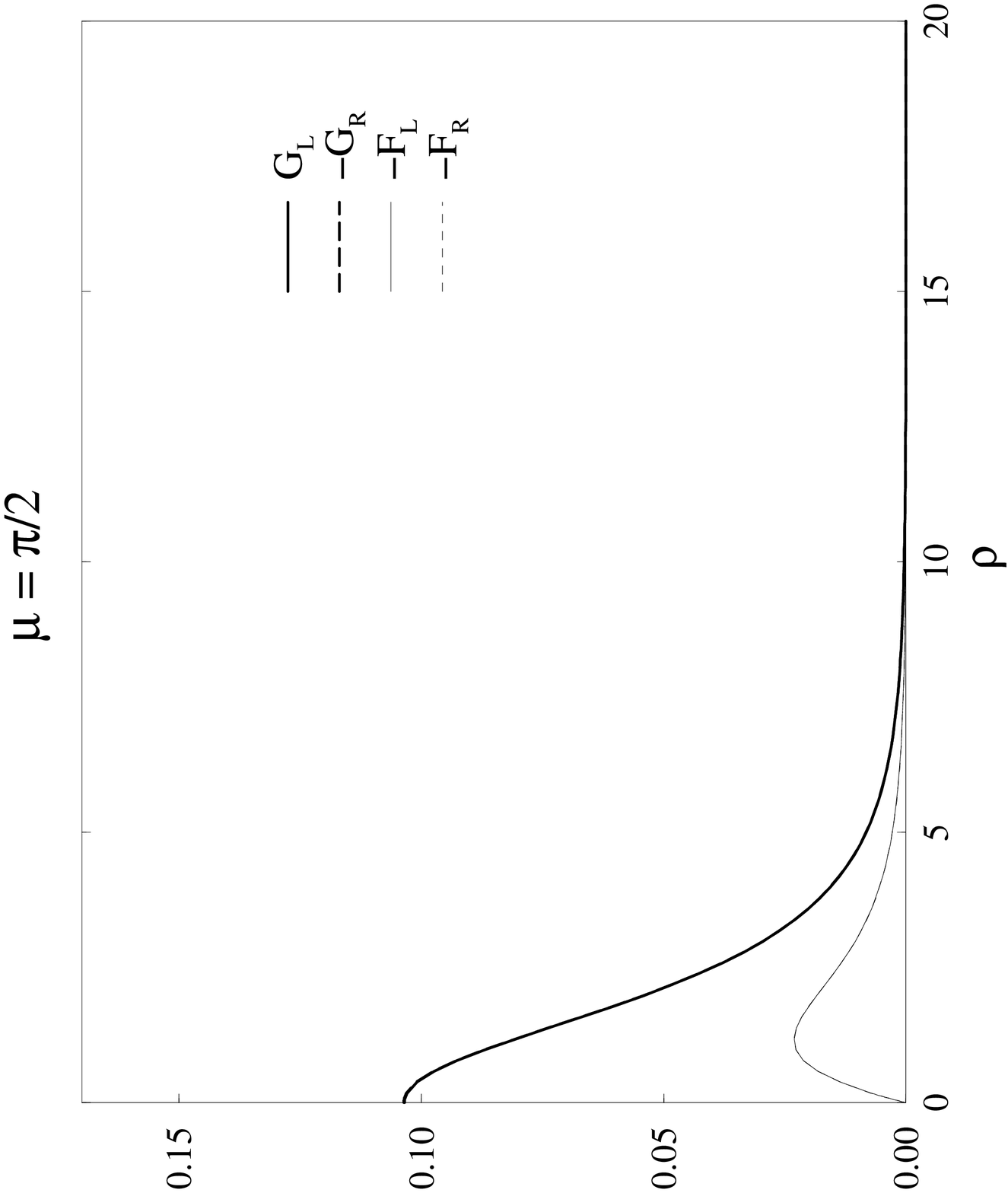}\\[0.2cm]

\caption{Fermion profile functions for $0\le\mu\le\pi/2$ and $\nu=0$.}
\label{Bildwindeinslepi}
\end{figure}
 
\begin{figure}
\rotpicsmall{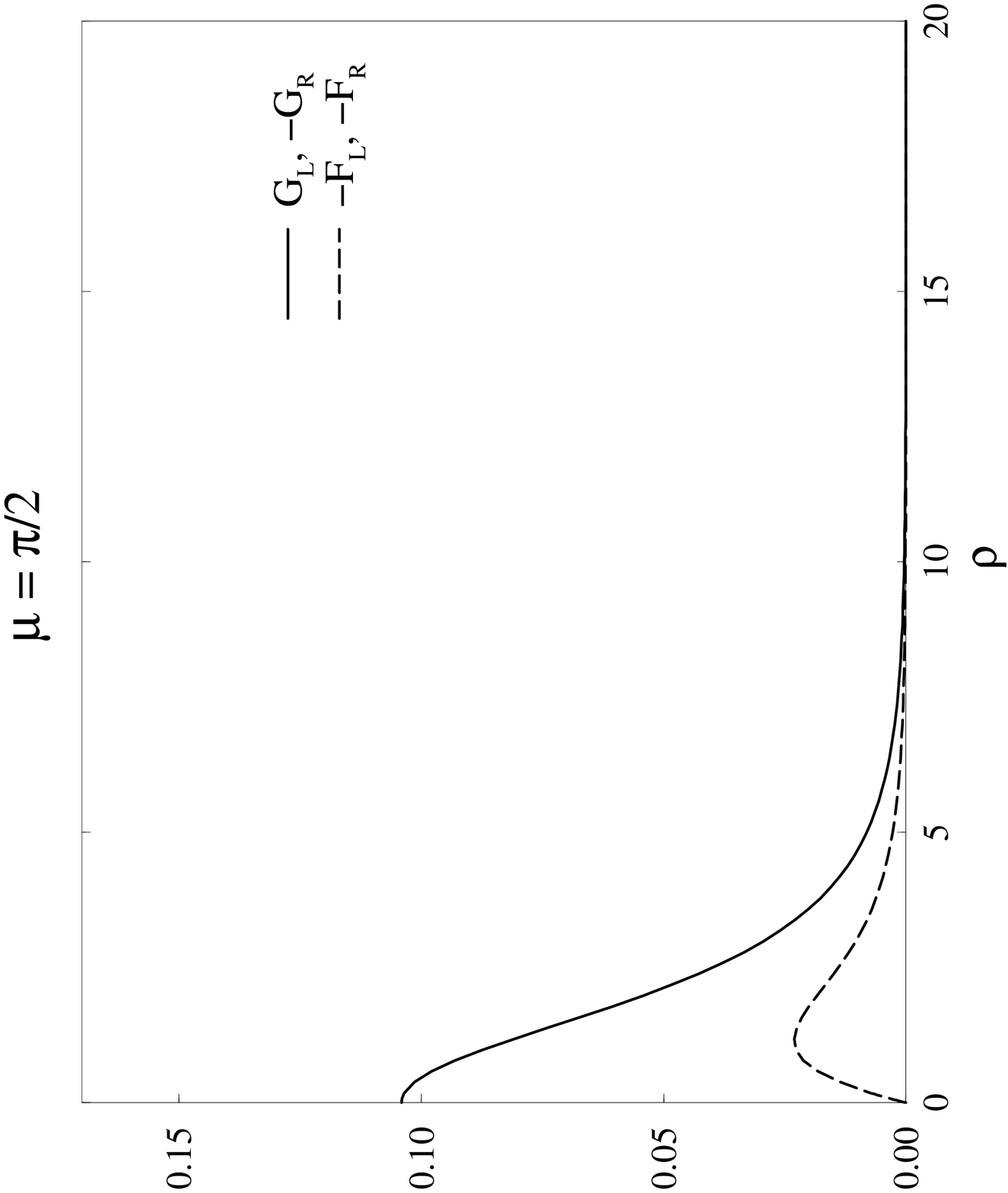}
\rotpicsmall{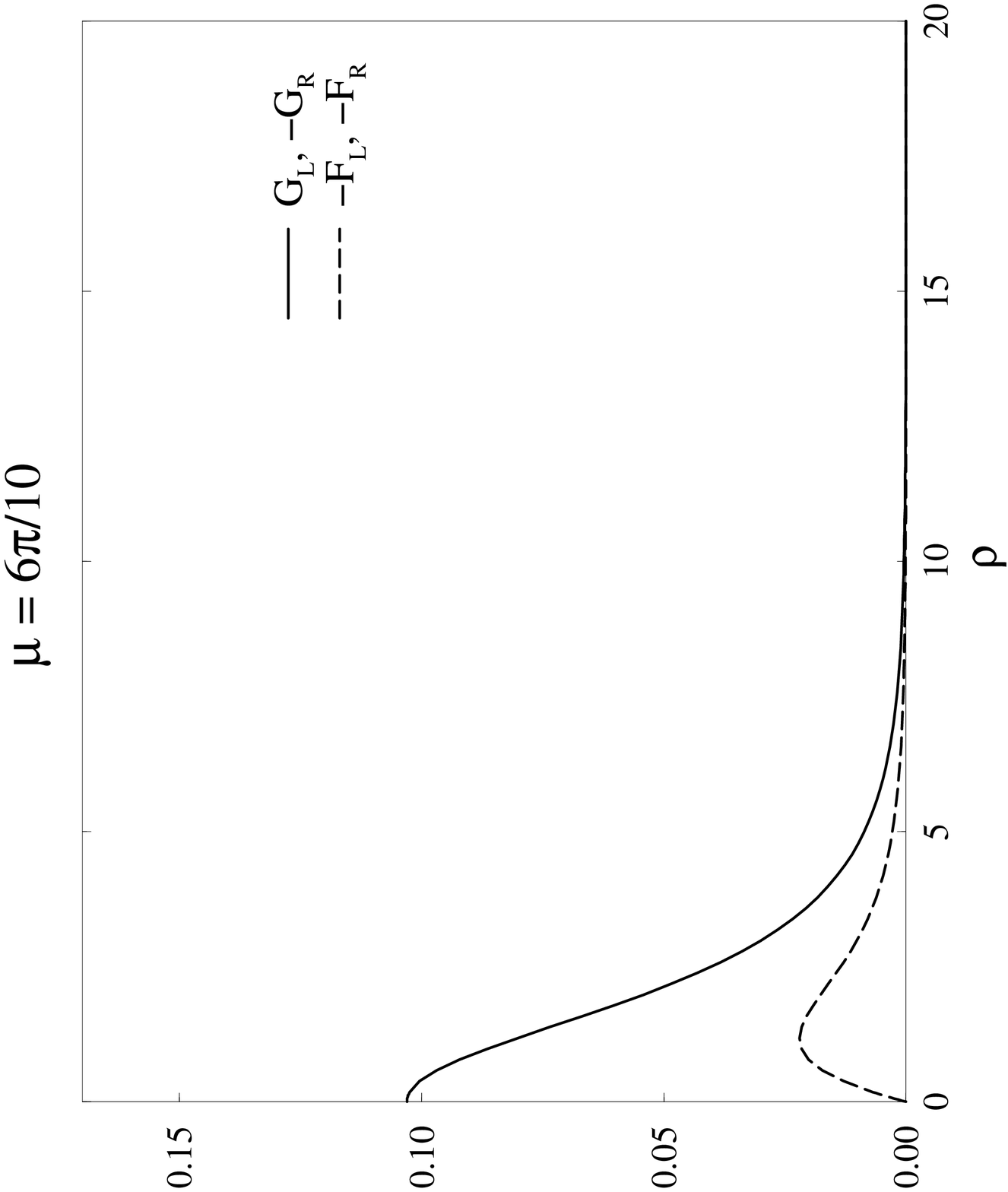}\\[0.2cm]

\rotpicsmall{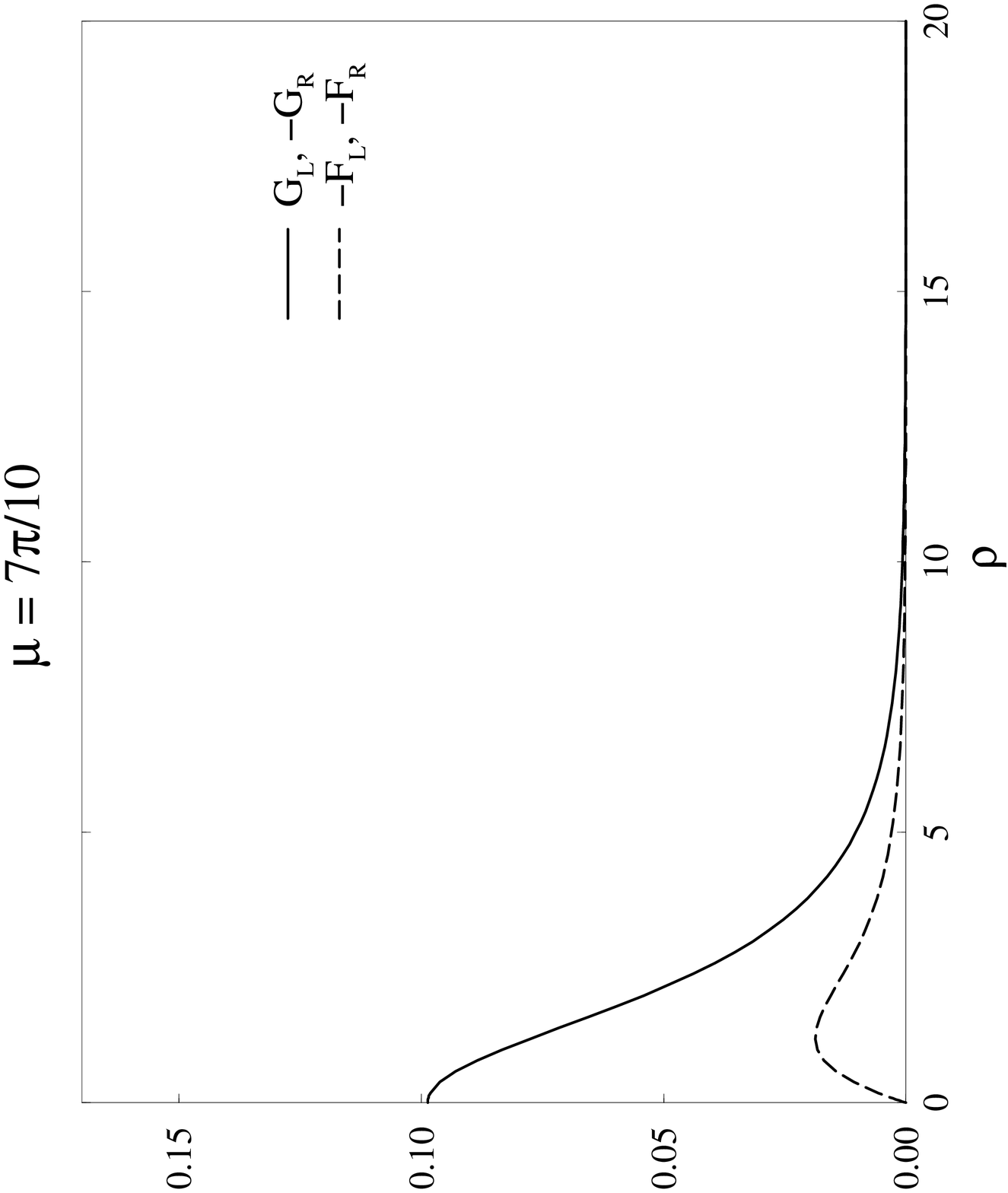}
\rotpicsmall{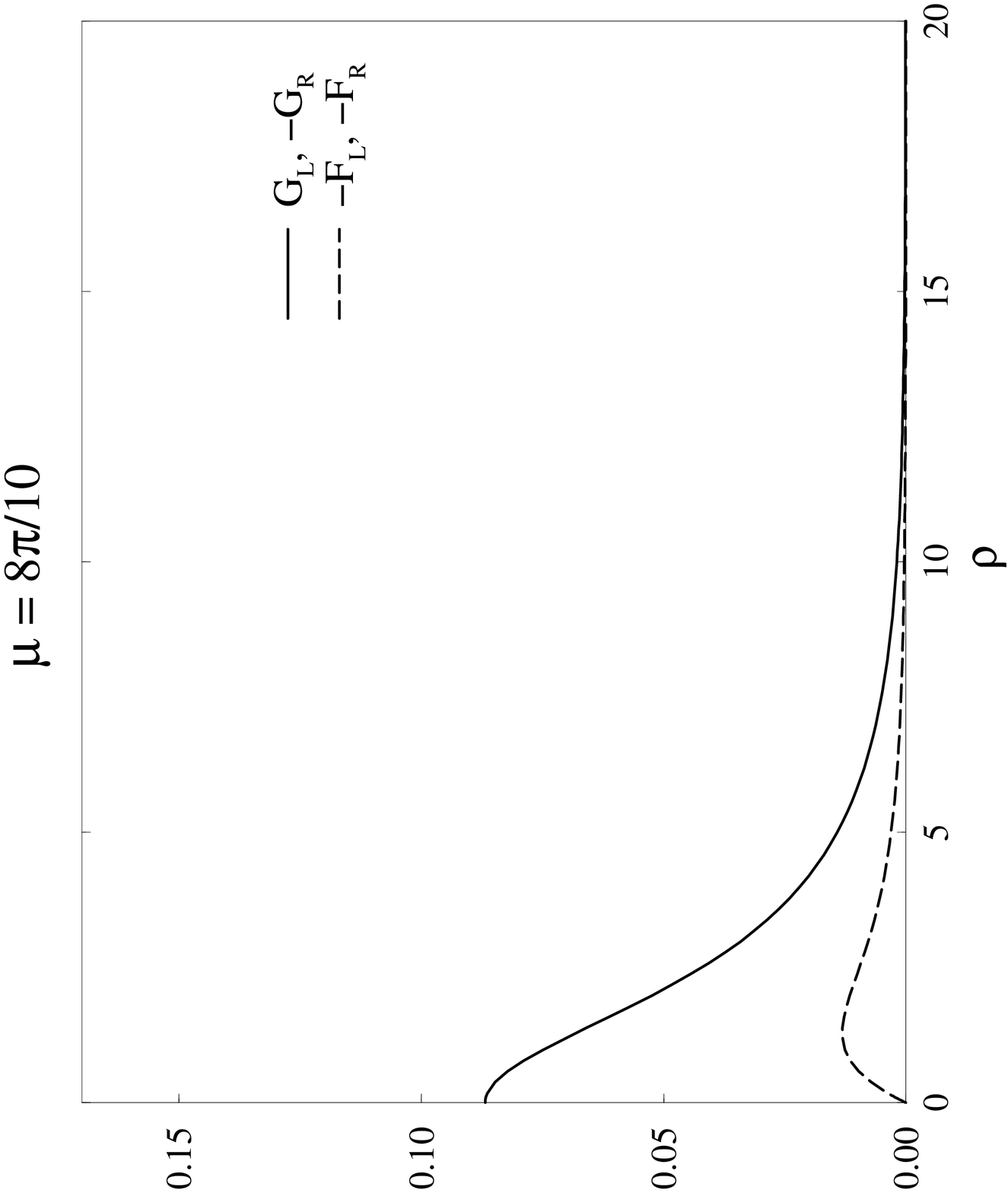}\\[0.2cm]

\rotpicsmall{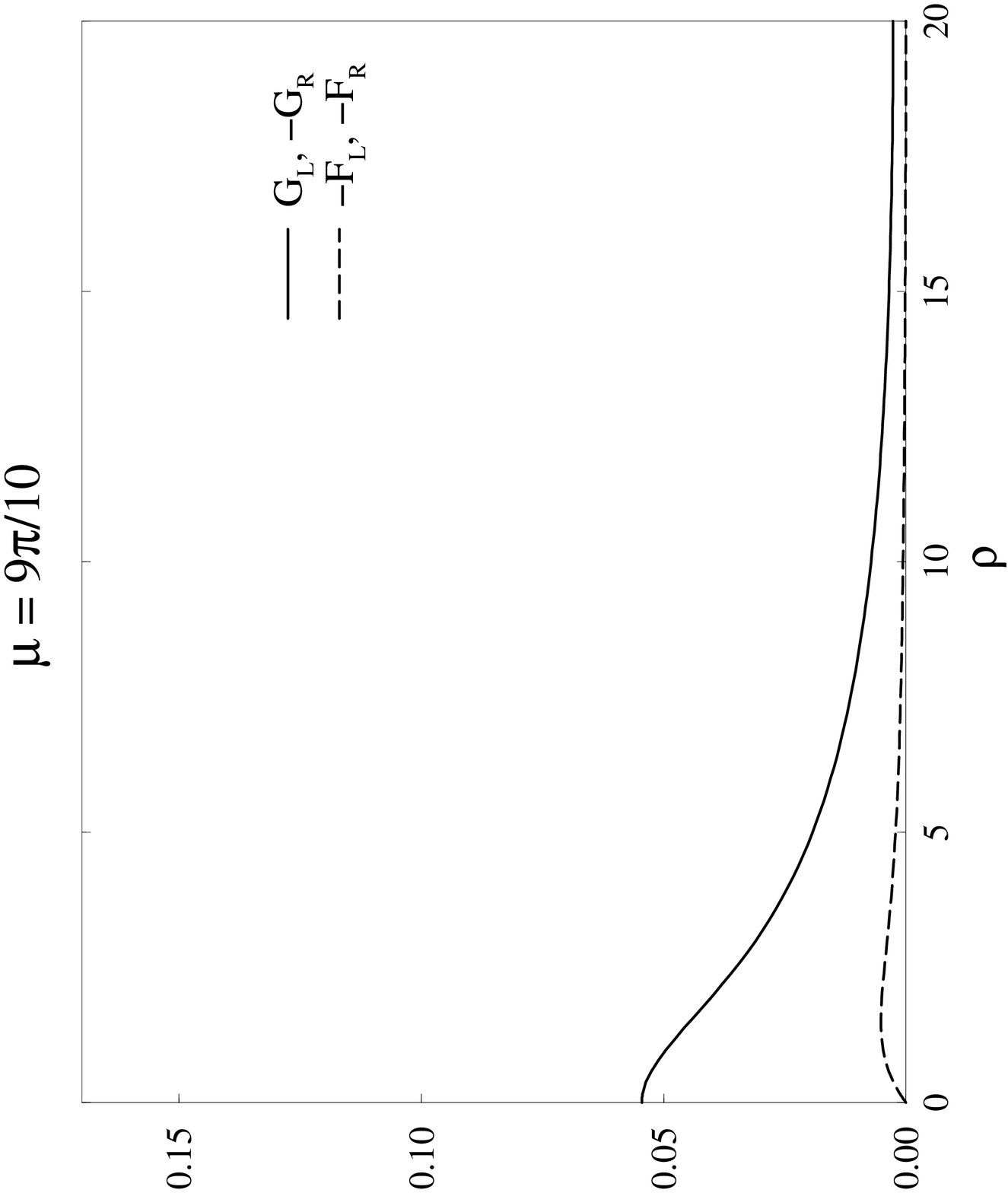}
\rotpicsmall{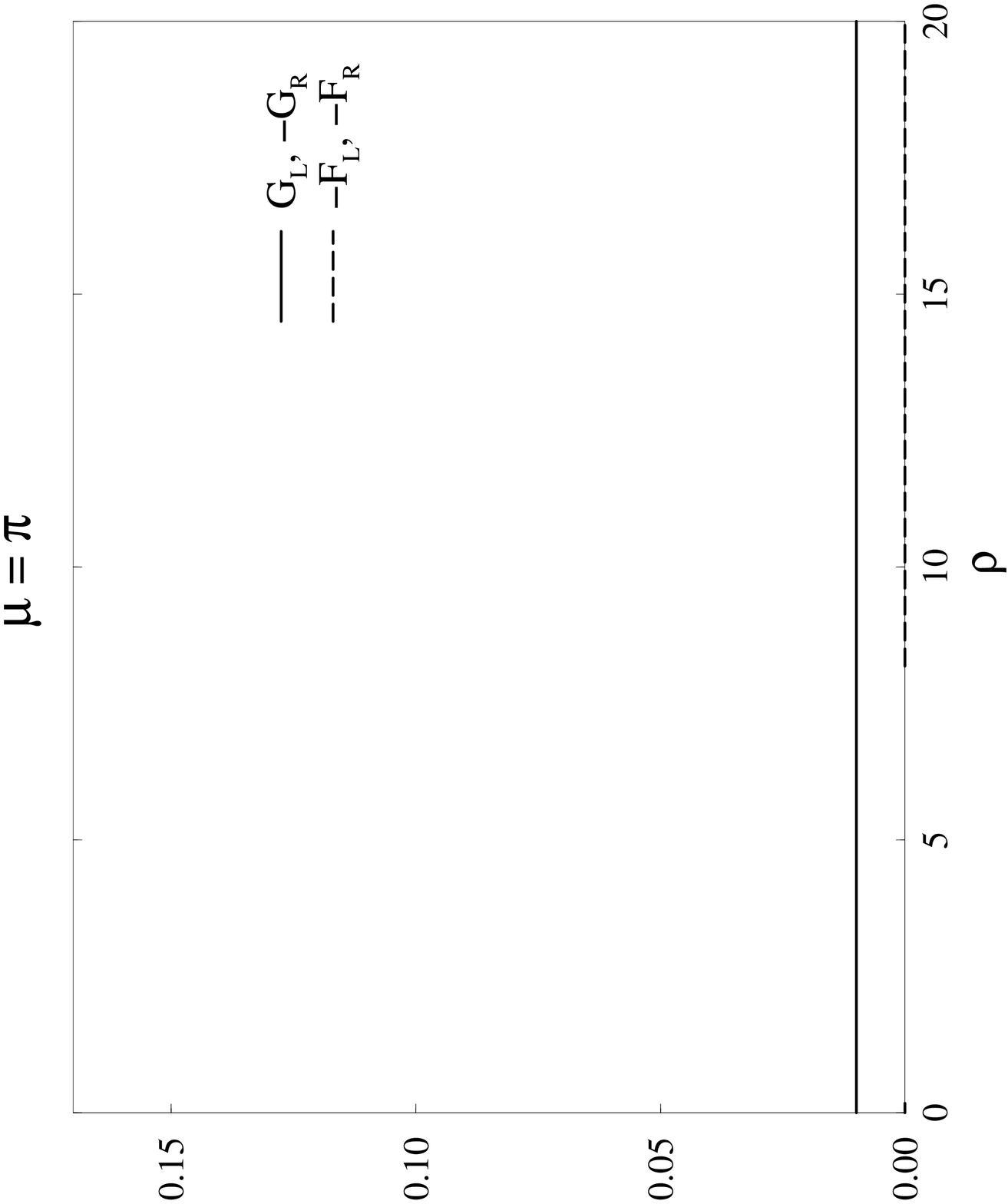}\\[0.2cm]

\caption{Fermion profile functions for $\pi/2\le\mu\le\pi$ and $\nu=0$.}
\label{Bildgepi}
\end{figure}
\par
For the numerical calculations the quantities $\rho$, $m$,
$E$, $F_i$ and $G_i$ are rescaled so that they become dimensionless
\beq                                       
\tilde \rho \equiv \rho \, M_W \, , \;\;
\tilde m    \equiv m \, M_W^{-1} \, , \;\;
\tilde E    \equiv E \, M_W^{-1}\, ,  \;\;
\tilde F_i  \equiv F_i \, M_W^{-3/2} \,, \;\;
\tilde G_i  \equiv G_i \, M_W^{-3/2}\,.
\label{rescaling}
\eeq
The tildes are dropped in the following.
The numerical results for the functions $G_L$, $G_R$, $F_L$ and $F_R$ are shown
in Figs. \ref{Bildwindeinslepi} and \ref{Bildgepi}, for the case of
equal masses $m=M_H=1$, and the corresponding energy eigenvalues in
Fig. \ref{Bildwindeinsemu} of the main text. The functions are calculated
for values of $\rho$ between $\rho_{\rm min}=10^{-3}$ and $\rho_{\rm max}=20$
and are normalized as follows
\beq                                       
||\Psi||^2 \equiv \int_0^{2\pi} {\rm d}\phi
\int_{\rho_{\rm min}}^{\rho_{\rm max}} \, {\rm d}\rho\,\rho\;
\Psi^\dagger(\rho,\phi) \Psi(\rho, \phi)  = 1 \, .
\label{normalization}
\eeq
This is then the reason that in Fig. \ref{Bildgepi}
the functions $G_L$ and $G_R$ for $\mu=\pi$ take on
constant, but nonzero, values. If the interval of normalization were
$[0,\infty)$, they would vanish identically.

In order to decide if the energy levels cross at $\mu=0$ or merely touch, we consider 
the continuity of the wavefunction. In fact,
the behaviour of $E(\mu)$ is determined completely
by demanding that the wavefunction varies continuously with $\mu$. 
To see this, write the ansatz (\ref{ansatz}) in a more general form
\bea
\Psi(\rho,\phi) &=& \phantom{+} i \, G_{L1}\, |L\uparrow d\ra
+ F_{L1}\, e^{i\phi} |L\downarrow d\ra \nonumber
+ i \, G_{L2}\, |L\downarrow u\ra + F_{L2}\, e^{-i\phi}|L \uparrow u \ra + 
\nonumber \\[0.1cm]
& & \phantom{i}G_{R1}\, |R\downarrow d\ra
+ i \, F_{R1}\, e^{-i\phi} |R\uparrow d\ra
+ G_{R2}\, |R\uparrow u\ra + i \, F_{R2}\, e^{i\phi} |R \downarrow u \ra \, .
\label{w1ansatz}
\eea
Solutions of the eigenvalue equation $H\,\Psi=E\,\Psi$ can be 
expressed in terms of the solutions $G_L$, $G_R$, $F_L$ and $F_R$
of (\ref{WindeinsGln}) for $\mu\geq0$ and $E\geq 0$ as follows
\begin{equation}
\begin{array}{l|cccccccc}
   & G_{L1} & G_{L2} & G_{R1} & G_{R2} & F_{L1} & F_{L2} & F_{R1} &
   F_{R2} \rule[-2ex]{0cm}{5ex}\\
\hline \hline
+\mu, +E & + \, G_{L} & - \, G_{L} & + \, G_{R} & + \,
G_{R} & + \, F_{L} & - \, F_{L} &
+ \, F_{R} & + \, F_{R}\rule[-2ex]{0cm}{5ex} \\
\hline
+\mu, -E & + \, G_{L} & + \, G_{L} & + \, G_{R} & - \,
G_{R} & - \, F_{L} & - \, F_{L} &
- \, F_{R} & + \, F_{R} \rule[-2ex]{0cm}{5ex}\\
\hline
-\mu, +E & + \, G_{L} & + \, G_{L} & + \, G_{R} & - \, G_{R} &
+ \, F_{L} & + \, F_{L}
& + \, F_{R} & - \, F_{R} \rule[-2ex]{0cm}{5ex}\\
\hline
-\mu, -E & + \, G_{L} & - \, G_{L} & + \, G_{R} & + \,
G_{R} & - \, F_{L} & + \, F_{L} &
- \, F_{R} & - \, F_{R}\rule[-2ex]{0cm}{5ex}
\end{array} \;\;\; .
\label{regeln}
\end{equation}
We now look at the over-all signs of these eight functions.
The signs of the four basic functions $G_L$, $G_R$, $F_L$
and $F_R$ can be read from Fig. (\ref{Bildwindeinslepi}).
The resulting signs of $G_{L1}$, $G_{L2}$, $G_{R1}$, $G_{R2}$,
$F_{L1}$, $F_{L2}$, $F_{R1}$ and $F_{R2}$
are given -- in this order -- by the following table:
\[
\begin{array}{c|ccc}
& & &\\
E>0 \hspace{.75em} & \hspace{.75em}(+,+,-,+,-,-,-,+) &    & (+,-,-,-,-,+,-,-)
                                                                \hspace{.5em}\\
& & &\\
E=0 \hspace{.75em}&                   & \begin{array}{c}
                           (+,-,-,-,0,0,0,0)\\
                           (+,+,-,+,0,0,0,0)
                           \end{array}
                                             &\\
& &  &\\
E<0 \hspace{.75em} & \hspace{.75em}(+,-,-,-,+,-,+,+) &     & (+,+,-,+,+,+,+,-)
                                                                \hspace{.5em}\\
& &  &\\
\hline
& &  &\\
& \mu<0 & \mu=0 & \mu>0
\end{array}
\]
If the wavefunction is traced from negative to positive values of $\mu$,
say, the relative signs of the eight functions cannot change abruptly.
From the sign table above, it then follows that the energy levels really cross.
For time-dependent background fields ($\mu=\mu(t)$, $\nu = 0$) this results   
in pair creation.
The role of particle production in anomalies has been emphasized in
\cite{NAG} and, evidently, pair production also plays a role in the
global gauge anomaly of Section 6 of this paper.

\section{Berry phase close to the $Z$-string}
\renewcommand{\theequation}{B.\arabic{equation}}
\setcounter{equation}{0}

\renewcommand{\thefigure}{B.\arabic{figure}}
\setcounter{figure}{0}

In this Appendix we calculate the Berry phase for loops on the $Z$-NCS that
are very close to the $Z$-string, using the original method of \cite{B}.

In that paper \cite{B}, Hamiltonians are considered which depend on three
external parameters $X_a$.
Discussed are loops in parameter space that are near the point ($X_a=0$)
at which the Hamiltonian has a degenerate eigenvalue.  If two states are
involved in this degeneracy, then the Berry phase can be calculated within
the subspace spanned by these two states (projection operator $\Pi$). 
Setting the degenerate eigenvalue to zero,
the Hamiltonian in the restricted space has the following form:
\beq
\Pi\, H \, \Pi = \half \mat{X_3 & X_1 -i\,X_2\\ X_1+i\,X_2 & -X_3} \, .
\label{gewuenschteform}
\eeq
It is shown in \cite{B} that the Berry phase corresponding to a simple
loop $X_a(\alpha)$, $\alpha \in [0,2\pi]$, is given by 
\beq
\gamma = \pm \,\frac{1}{2}\, \Omega\, ,
\eeq
where $\Omega$ is the solid angle under which the loop is seen from the point 
of degeneracy ($X_a=0$) and the sign depends on the state considered.
If the loop lies in a plane through the origin, then there are 
two possibilities: either the origin is encircled once by the loop and
$\gamma =\pm \pi$ or the origin lies outside the loop and $\gamma=0$.

These considerations apply to our case, with the $Z$-NCS fields
as the external parameters. In the background of the $Z$-string ($\mu=0$,
$\nu=0$), we have indeed two degenerate fermion zero modes
$|\Psi^\pm (0,0)\ra \,$, called $\Psi_1$ and $\Psi_2$ in Appendix A.
We now parametrize the Dirac Hamiltonians (\ref{Hamiltonian}) 
by $\mu_0 \equiv \arccos c$  and $\alpha \equiv 2\pi t/T \in [0,2\pi]$,
for the contour $\cos\mu\cos\nu=c$ on the $Z$-NCS.
See also (\ref{Hmu0t}) in the main text. Since we only need curves that are 
close to $\mu_0=0$, we can expand $H(\mu_0,\alpha)$ to first order in $\mu_0$.
A straightforward calculation gives then
\beq                                         
\la \Psi^\pm |\, H \,| \Psi^\pm \ra =          
 \mu_0 \, E^{(1)} \, \mat{\cos\alpha & i\sin\alpha
 \\ -i \sin\alpha & -\cos\alpha}\; + \, {\rm O}(\mu_0^2) \, ,
\label{Hmatrix}
\eeq                                       
with $E^{(1)}$ a non-vanishing constant, which is expressed
in terms of the bosonic and fermionic profile functions of the $Z$-string by
\beq                                                         
E^{(1)} = 4\,\pi \,M_W\,\int_0^\infty {\rm d}\rho \, \rho\,
          \left( \frac{f}{\rho} \, G_L^{\, 2} -
          2\, m\, G_L\, G_R
          \right) \,,
\label{E1}                                                         
\eeq
where all quantities inside the integral are dimensionless as in
(\ref{rescaling}, \ref{normalization}).
The same result holds for the $Z$-NCS\p, with the factor $m$ in the second
term of the integrand of $E^{(1)}$ replaced by  $m\, h$.
Comparing (\ref{Hmatrix}) with the general form (\ref{gewuenschteform}),
we have 
\beq
X_1(\alpha)=0 \,,\quad  
X_2(\alpha)= - 2\, \mu_0 E^{(1)}\sin\alpha \, , \quad 
X_3(\alpha)=2\, \mu_0 E^{(1)}\cos\alpha \, . 
\eeq
This curve $X_a(\alpha)$ lies in a plane through
the origin and for $\mu_0$ fixed encircles it once as $\alpha$ runs from $0$ to $2\pi$,
so that the Berry phase factor is $e^{i\gamma}=-1$.
More general closed curves C near the $Z$-string give the result
(\ref{gammaC}) in the main text.

\section{Berry phase on the $Z^2$-NCS\p}
\renewcommand{\theequation}{C.\arabic{equation}}
\setcounter{equation}{0}

\renewcommand{\thefigure}{C.\arabic{figure}}
\setcounter{figure}{0}

In this Appendix we calculate for isodoublet fermions the spectral flow and 
the Berry phase over a \ncs ~with winding number $n=2$.

\begin{figure}
  \parbox{7.5cm}{
    \rotpicsmall{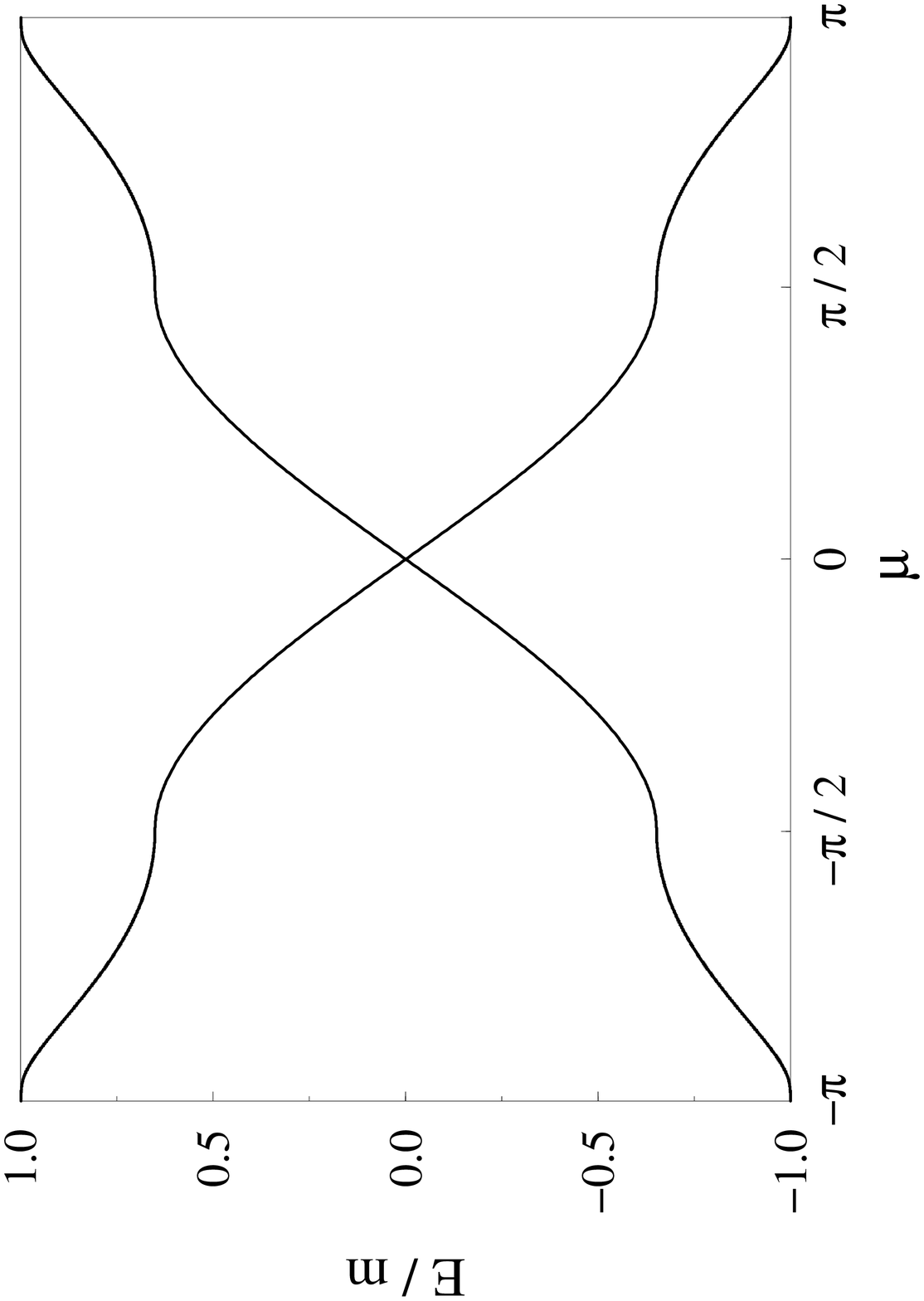}
    \caption{\small
      Energies for $n=2$, $\nu=0$. 
      \label{Bildwindzweiemu}
    }
  }
  \hfill
  \parbox{7.5cm}{
    \rotpicsmall{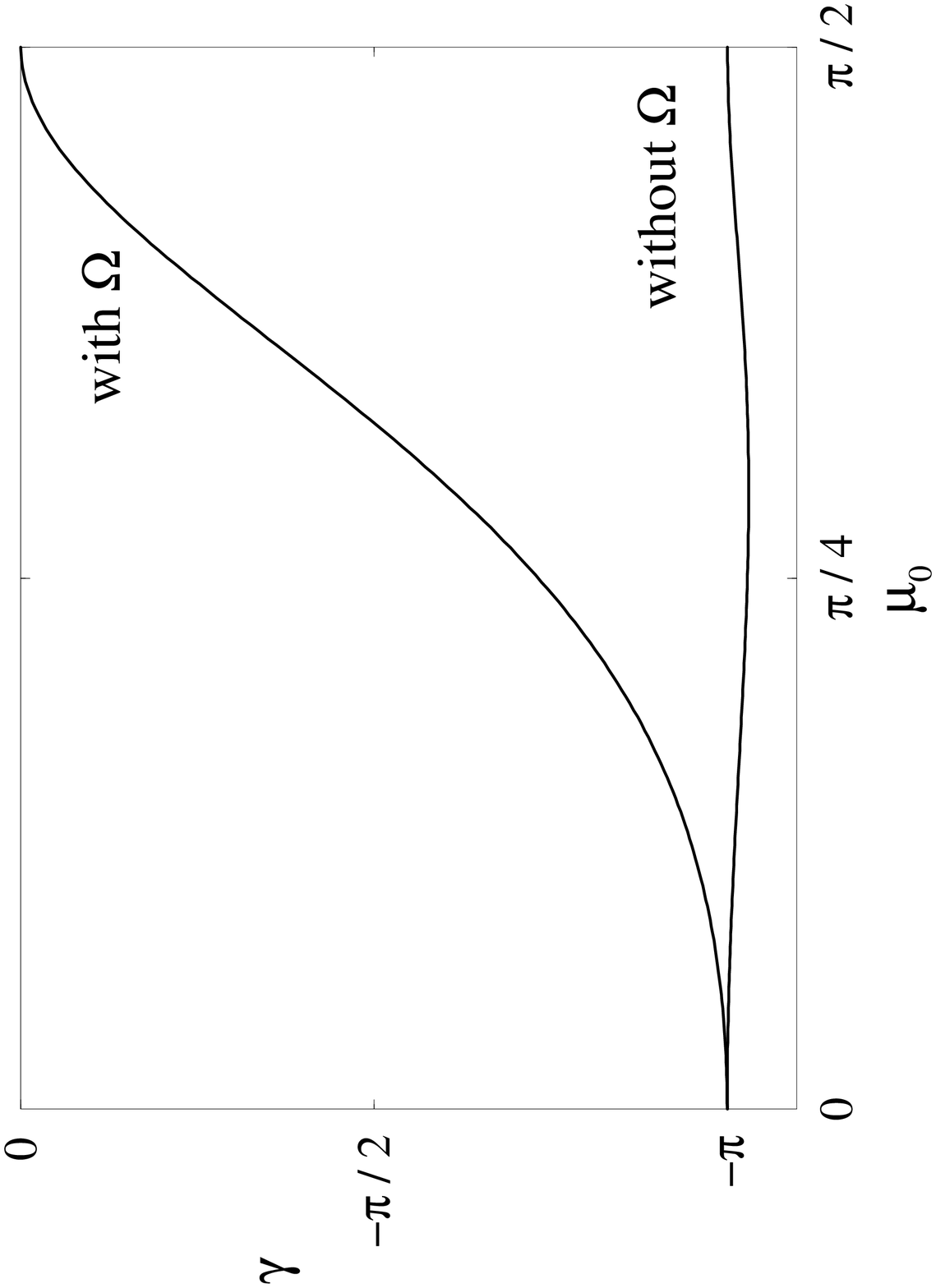}
    \caption{\small
      Berry phase for $n=2$.      
      \label{Berrybild}
    }
  }
\end{figure}

The $Z$-NCS\p ~(\ref{felddef}, \ref{matrizen}) can be generalized by
replacing $\phi$ with $n\, \phi$, so that the matrix function
$U$ has winding number $n\in \Z$. The resulting set of bosonic field configurations
is called the $Z^n$-NCS\p.  The axial symmetry of the fermion fields
is then generated by
\beq
K_3^{(n)}\equiv -i\,\partial_\phi + \half\,\Sigma_3 +\frac{n}{2}\,\tau_3\, (P_L-P_R)\,.
\eeq
As an example, we consider the case $n=2$. 
The Dirac eigenvalues for the $\nu=0$ slice of the $Z^2$-NCS\p ~are now
doubly degenerate, corresponding to the $K_3^{(2)}$-eigenvalues $\pm \half$,
and have no bifurcation at $\mu=\pm\,\pi/2$, see Fig. \ref{Bildwindzweiemu}.
At $\mu=0$ there is a four-fold degeneracy, of course. It can be
shown that the levels there really cross, by the same type of argument as 
used in Appendix A.

\begin{figure}
\centerline{\mbox{
\parbox[t]{7cm}{
\unitlength1cm

\begin{picture}(7,5)
\put(-3,7.5){\includegraphics{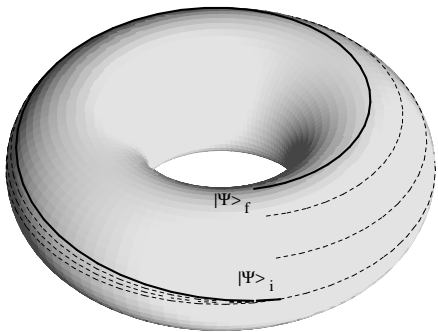}}
\end{picture}
}}}
\caption{Torus bundle for a slice of the $Z^2$-NCS\p, with different Berry phases.
\label{torusbild}}
\end{figure}

As regards the Berry phase $\gamma$, there is an important
difference compared to the case of winding number $n=1$.
The configuration space of the Dirac field has to be restricted to an
eigenspace of $K_3^{(2)}$ with non-degenerate energies. However,
the relevant $K_3^{(2)}$ eigenstates are {\em not\/} real
and therefore the Berry
phase factor $e^{i\gamma}$ is {\em not\/} restricted to the values $\pm 1$.
Indeed, a numerical calculation based on (\ref{Psisolution})
shows that $e^{i\gamma}$ depends on the loop of Hamiltonians chosen
and is in general complex. What is more, the Berry phase depends on
whether the matrix $\Omega$ in (\ref{felddef}, \ref{matrizen})
is included or not.  This is shown in Fig. \ref{Berrybild}, 
where for both cases the Berry phase of the $K_3^{(2)}=+\half$ eigenstate is plotted against 
$\mu_0$, which corresponds to
the point where the $\cos\mu\cos\nu=c$ contour intersects the positive
$\mu$-axis ($\mu_0 = \arccos c$).  
Interpolation between these two alternatives (with or without $\Omega$) is
also possible, so that {\em a priori\/} any intermediate value of the Berry phase $\gamma$
can be obtained.
For the $Z^2$-NCS\p ~with $\mu_0$ fixed and $\alpha\in[0,2\pi]$
there is, in fact, a complex torus bundle (Fig. \ref{torusbild}).
Still, the true $Z^2$-NCS\p ~does involve the matrix $\Omega$ and
the Berry phase actually disappears as the loop is pulled towards the vacuum
($\mu_0 \rightarrow \pi/2$ in Fig. \ref{Berrybild}).
Hence, we expect that the vacuum loop of the $Z^2$-NCD,
given by (\ref{ZNCDfields}, \ref{ZNCDmatrices}) with $\eta=\pi/2$ and 
$\phi$ replaced by $2\, \phi$, has a trivial Berry phase factor $+1$, corresponding to a
cylinder bundle over the $n=2$ gauge orbit (Fig. \ref{moebiusbild}).


\section{Gauss' law and the global anomaly}
\renewcommand{\theequation}{D.\arabic{equation}}
\setcounter{equation}{0}

\renewcommand{\thefigure}{D.\arabic{figure}}
\setcounter{figure}{0}

In this Appendix we give a brief discussion of the
physical state condition (Gauss' law) and the problem that arises
due to the global gauge anomaly.

Consider the hamiltonian quantization, in the temporal gauge,
of a $3+1$ dimensional \YMth ~with chiral fermions (the Higgs field is not
important for the present discussion). In the Schr\"{o}dinger
representation the states $\Psi$ are functionals of the gauge fields
$W_m(\mathbf{x})$ and infinite component column vectors in the fermionic Hilbert
space. Physical states $\Psi_{\rm phys}$ obey the condition \cite{J}
\beq
\phantom{1}^{\rm F} \mathcal{G}_{\,U} \, \Psi_{\rm phys}\, \left[\,W_m^{\,U}\,\right] =
\Psi_{\rm phys}\, \left[\,W_m\phantom{^U}\!\!\right]    \, ,
\label{physcond}
\eeq
with $W_m^{\,U} \equiv U \, (W_m + \partial_m) \, U^{-1}$ the
gauge transform of the field $W_m$ and
$\!\phantom{1}^{\rm F} \mathcal{G}_{\, U}$
the cor\-res\-pon\-ding unitary fermionic transformation.
The condition (\ref{physcond}) incorporates the physical requirement of
gauge invariance and its infinitesimal version, for
$U(\mathbf{x}) = \id + \theta(\mathbf{x})$ with $\theta(\mathbf{x})$ in the Lie algebra
of the gauge group $G$, gives the so-called non-Abelian Gauss' law.
The gauge transformation function $U(\mathbf{x})\in G$ is considered to be topologically
trivial, so that there are no additional phase factors in
(\ref{physcond}). As emphasized in \cite{NAG},
this physical state condition (\ref{physcond})
respects the real structure of the second-quantized Hilbert bundle.

The global gauge anomaly shows up in the following way.
The starting point is the existence of a loop of
3-dimensional gauge transformations $U(\omega)\in G$,
with $\omega \in [0,2\pi]$ and $U(0)=U(2\pi)=\id$,
that gives a Berry phase factor $-1$
for one particular state $\bar{\Psi}_{\rm phys}$ (in this paper, the vacuum state).
This non-trivial Berry phase factor may come from having an \emph{odd} number of pairs of fermionic levels crossing at $E=0$.
The result
\beq
  \phantom{1}^{\rm F} \mathcal{G}_{\,U(2\pi)} \,
  \bar{\Psi}_{\rm phys}\, \left[\,W_m^{\,U(2\pi)}\,\right]       =
- \phantom{1}^{\rm F} \mathcal{G}_{\,U(0)} \,
  \bar{\Psi}_{\rm phys}\, \left[\,W_m^{\,U(0)}\,\right]       =
- \, \bar{\Psi}_{\rm phys}\, \left[\,W_m\phantom{^U}\!\!\right]    
\label{minusGF}
\eeq
is, however, incompatible with (\ref{physcond}).
In other words, Gauss' law cannot be implemented continuously over 
the \emph{whole} of the space of Yang-Mills connections and the theory 
is said to have a global anomaly. 
See \cite{NAG,J} for further details and discussion.
Here, we only remark that there may be different types of gauge
transformation loops $U(\omega)$ which produce a factor $-1$ in
(\ref{minusGF}), see in particular (\ref{Uvac}) and (\ref{UWitten}) in the main text.

\end{appendix}


\newpage
\begin{thebibliography}{99}
\bibitem{KO} F. Klinkhamer and P. Olesen, Nucl. Phys. {\bf B 422} (1994), 227.
\bibitem{NO} H. Nielsen and P. Olesen, Nucl. Phys. {\bf B 61} (1973), 45.
\bibitem{N}  Y. Nambu, Nucl. Phys. {\bf B 130} (1977), 505.
\bibitem{B}  M. Berry, Proc. R. Soc. Lond. {\bf A 392} (1984), 45.
\bibitem{W}  E. Witten, Phys. Lett. {\bf B 117} (1982), 324.
\bibitem{NAG}P. Nelson and L. Alvarez-Gaum{\'e},
             Comm. Math. Phys. {\bf 99} (1985), 103.
\bibitem{JR} R. Jackiw and P. Rossi, Nucl. Phys. {\bf B 190} (1981), 681.
\bibitem{EP} M. Earnshaw and W. Perkins, Phys. Lett. {\bf B 328} (1994), 337.
\bibitem{C}  S. Coleman, {\it Aspects of symmetry\/}
             (Cambridge U. P., Cambridge, 1985), Chapter 7. 
\bibitem{JRe}R. Jackiw  and C. Rebbi, Phys. Rev. {\bf D 16} (1977), 1052.
\bibitem{BPST} A. Belavin, A. Polyakov, A. Schwartz and Yu. Tyupkin, 
               Phys. Lett. {\bf B 59} (1975),~85.
\bibitem{K}  F. Klinkhamer, Nucl. Phys. {\bf B 410} (1993), 343.
\bibitem{J}  R. Jackiw, in B. DeWitt and R. Stora (eds.), {\em Relativity,
             Groups and Topology II\/} (North-Holland, Amsterdam, 1984).
\bibitem{Bar}W. Bardeen, Phys. Rev. {\bf 184} (1969), 1848.
\bibitem{O}  S. Okubo, C. Geng, R. Marshak and Z. Zhao,
             Phys. Rev. {\bf D 36} (1987), 3268.
\bibitem{VS} H. de Vega and F. Schaposnik, Phys. Rev. {\bf D 14} (1976), 1100.
\end{thebibliography}
\end{document}